\documentclass[aps,twocolumn,floatfix,superscriptaddress,showpacs,prb]{revtex4-1}

\usepackage{epsf}
\usepackage{epsfig}
\usepackage{graphicx}
\usepackage{dcolumn}
\usepackage{braket}
\usepackage{bm}
\usepackage{amsfonts}
\usepackage{amsmath}
\usepackage{amssymb}
\usepackage{color,soul}
\usepackage{natbib}

\usepackage{txfonts}

\definecolor{light-gray}{gray}{0.5}

\newcommand{\tf}[1]{\mathrm{#1}}
\newcommand{\dd}[2]{\frac{\partial {#1}}{\partial {#2}}}

\newcommand{\tr}[1]{\underset{\mathrm{#1}}{\mathrm{tr}}}

\begin{document}

\title{Real-time diagrammatic approach to current-induced forces: Application to quantum-dot based nanomotors}

\author{Hern\'an L. Calvo}
\affiliation{Instituto de F\'isica Enrique Gaviola (CONICET) and FaMAF, Universidad Nacional de C\'ordoba, Argentina}
\affiliation{Departamento de F\'isica, Universidad Nacional de R\'io Cuarto, Ruta 36, Km 601, 5800 R\'io Cuarto, Argentina}

\author{Federico D. Ribetto}
\affiliation{Instituto de F\'isica Enrique Gaviola (CONICET) and FaMAF, Universidad Nacional de C\'ordoba, Argentina}
\affiliation{Departamento de F\'isica, Universidad Nacional de R\'io Cuarto, Ruta 36, Km 601, 5800 R\'io Cuarto, Argentina}

\author{Ra\'ul A. Bustos-Mar\'un}
\affiliation{Instituto de F\'isica Enrique Gaviola (CONICET) and FaMAF, Universidad Nacional de C\'ordoba, Argentina}
\affiliation{Facultad de Ciencias Qu\'imicas, Universidad Nacional de C\'ordoba, Argentina}

\begin{abstract}
During the last years there has been an increasing excitement in nanomotors and particularly in current-driven nanomotors.
Despite the broad variety of stimulating results found, the regime of strong Coulomb interactions has not been fully explored for this application.
Here we consider nanoelectromechanical devices composed by a set of coupled quantum dots interacting with mechanical degrees of freedom taken in the adiabatic limit and weakly coupled to electronic reservoirs. We use a real-time diagrammatic approach to derive general expressions for the current-induced forces, friction coefficients, and zero-frequency force noise in the Coulomb blockade regime of transport. We prove our expressions accomplish with Onsager's reciprocity relations and the fluctuation-dissipation theorem for the energy dissipation of the mechanical modes. The obtained results are illustrated in a nanomotor consisting of a double quantum dot capacitively coupled to some rotating charges. We analyze the dynamics and performance of the motor as function of the applied voltage and loading force for trajectories encircling different triple points in the charge stability diagram. 
\end{abstract}


\maketitle

\section{Introduction} 

Not so long ago the scientific community wondered if current-induced forces (CIFs) in nanoscale devices 
could be used for something else other than heating and damaging the conductors.~\cite{diventra2004,dundas2009}
Few years later, the interest rapidly evolved towards the design and control of efficient nanomotors powered by direct currents.~\cite{bailey2008,wang2008,dundas2009,bode2011,bustos2013,fernandez2015,arrachea2015,celestino2016,ludovico2016,fernandez2017}
This last was also fueled by recent seminal experiments.~\cite{fennimore2003,barreiro2008,kudernac2011,tierney2011,lotze2012,kim2014}
The fast development of the topic is surely a consequence of the great interest it arouses. This is 
understandable considering that macroscopic engines have played a major role in the development of modern 
civilization and that biological nanomotors make complex life possible as we know it.~\cite{goel2008,guix2014}
These facts naturally awake the imagination towards the uncountable applications where the 
research could lead us one day. However, the development of efficient and reliable current-induced nanomotors 
is still an open challenge in current nanoscience and nanotechnology.

Recent theoretical works on the topic have shed light into the intrinsic mechanisms of the CIFs and its application to the 
development of current-driven nanomotors. For example, the origin of the nonconservative part of the CIF~\cite{todorov2011} and 
its sharp activation with bias voltages.~\cite{dundas2012} In general nonequilibrium conditions, it was shown that the CIF does not 
only contain a frictional term, but also a Lorentz-like term associated with a Berry-phase contribution.~\cite{bode2011} In 
molecular junctions, the CIF can induce a renormalization of the vibrational modes coupled to the molecule, thus affecting the 
structure and stability of the electronic device.~\cite{bai2016} The application of CIFs in nanomotors allowed for the establishment 
of a fundamental relation with the concept of adiabatic quantum pumping.~\cite{bustos2013} Indeed, this relation leads to the term 
``adiabatic quantum motor,'' and applies when the mechanical degrees of freedom are slow compared to the electronic time scales and 
can be treated as classical. In such devices, efforts were made in understanding the role of decoherence,~\cite{fernandez2015} together 
with the interplay between conservative forces, nonconservative ones, and dissipation in the motor dynamics.~\cite{fernandez2017} 
Moreover, it was predicted that for non-linear stochastic dynamics, the force fluctuations tend to enhance the pumping mechanism under 
resonant conditions.~\cite{perroni2014} In the context of ac-driven quantum systems, a generalized thermoelectric framework was derived 
to connect different response coefficients through Onsager's reciprocity relations.~\cite{ludovico2016} Applied to adiabatic quantum 
motors, for example, this allows one to relate the work done by the CIF with charge and heat pumped currents. Similar Onsager's relations 
were used to derive mutual electron-phonon drag effects through coherent molecular conductors and to relate them with both quantum pumping 
and CIFs.~\cite{liu2016}

Most of the above mentioned works deal with systems where the electron-electron interaction can be either neglected or treated on a mean-field level. 
However, research on CIFs is not restricted to this parameter range. Some related works are based on the Coulomb blockade regime of transport, characterized by a dominant electronic repulsion and a weak coupling to the electrodes. Examples are molecular rotary motors driven by electron tunneling,~\cite{wang2008,croy2012,celestino2016} where the force is exerted by an electrostatic field subtended between the leads; and quantum shuttles,~\cite{gorelik1998,novotny2003,fedorets2004} where a movable island transfers the electronic charges between source and drain leads.
Nonequilibrium Green's function methods, for example, can be used to include electron-electron interactions, though this is usually done perturbatively and its application can be cumbersome.~\cite{haug2008} Other techniques usually rely on the self-consistent time integration of an effective rate equation including both electronic and mechanical degrees of freedom. However, the separation between their time-scales is either not exploited or taken into account through \textit{ad-hoc} assumptions.

In view of this, it results desirable to explore adiabatic quantum motors through appropriate formalisms to include the strong Coulomb interaction exactly and, on the same time, able to exploit the separation between different time-scales. To give a complete understanding of the interplay between these degrees of freedom, it would be also important to include current-induced dissipation of the mechanical energy as well as current-induced noise in the forces.

In this work, we use a real-time diagrammatic approach~\cite{splettstoesser2006,riwar2010,calvo2012,haupt2013,riwar2013,juergens2013} 
to derive general expressions for the CIFs, friction coefficients, and random 
fluctuations of forces in many-body systems consisting of coupled quantum dots interacting locally with slow classical 
degrees of freedom (see Fig.~\ref{fig:fig1}). Taking advantage of the different time scales of the processes involved, the expressions derived here do not require the full integration of the time-dependent Liouville-von Neumann equation for the reduced density matrix of the local system, as transient effects can be disregarded. Instead, they are naturally obtained from a perturbative treatment in the oscillation frequency of the mechanical modes. The diagrammatic theory employed here provides a rigorous formal tool, derived from first principles, that allows one to clearly control the level of approximation in both the tunnel coupling and the modulation frequency.~\cite{splettstoesser2006} 
Although we restrict ourselves to leading order in the weak coupling to the leads and assume an adiabatic approximation for the classical mechanical degrees of freedom, the found expressions can be formally extended to higher orders in the adiabatic expansion and/or in the tunnel coupling.~\cite{splettstoesser2006,cavaliere2009}
It should be mentioned that there is a precedent of the application of the real-time diagrammatic approach to CIFs, done in Ref.~\onlinecite{holzbecher2014}. There, the authors numerically evaluated the work per cycle done by the CIFs and its connection with the pumped charge within the linear bias regime. This was done particularly in a potential nanomotor (or quantum pump) based on a double quantum dot and motivated by recent experiments on a carbon nanotube based mechanical resonator.~\cite{benyamini2014} In the present work, we extend those formulas to general quantum dot systems and include dissipation and force noise. In addition, we formally prove the Onsager's reciprocity relations connecting the CIFs to the tunnel currents as well as the fluctuation dissipation theorem for the force. We also explicitly treat the dynamical problem of the mechanical modes as well as the performance of the nanomotors in terms of the thermodynamic efficiency and the output power.

The paper is organized as follows. In section~\ref{sec:model} we present the general model that 
describes the type of system treated in this work. In section~\ref{sec:method} we briefly overview the 
real-time diagrammatic approach and then we give the general expressions for the CIFs
and current-induced friction coefficients. In this section we also derive within 
this formalism the Onsager's reciprocity relation for the charge currents and the CIFs and then we prove 
the fluctuation-dissipation relation between the force correlation function and the current-induced friction. In 
section~\ref{sec:dqd} we illustrate the role of these expressions in a double quantum dot based nanomotor, where each one of the dots interact with a charged rotor. We analyze its dynamics and performance in terms of the applied bias, loading force, and other parameters of the system. Finally, in section~\ref{sec:conclusions} we summarize the main results.

\section{Model and formalism}
\label{sec:model}

\subsection{General model}

\textit{Hamiltonian -} We consider quantum dot systems (from hereon the \textit{local system}) in which both 
electronic and mechanical degrees of freedom are present and coupled to each other. Such a 
local system is represented through the Hamiltonian
\begin{equation}
\hat{H}_\tf{sys} = \hat{H}_\tf{el}(\hat{\bm{X}})+\frac{\hat{\bm{P}}^2}{2m}+ U(\hat{\bm{X}},t), \label{eq:Hsys}
\end{equation}
where $\hat{\bm{X}} = (\hat{X}_1,...,\hat{X}_N)$ is the vector of mechanical coordinates and 
$\hat{\bm{P}} = (\hat{P}_1,...,\hat{P}_N)$ collects their associated momenta. $m$ is the effective mass related to 
$\hat{\bm{X}}$ and $U$ represents some external potential which might be present. The time dependence on $U$ emphasizes 
the fact that some external agent can exert work on the system.
The Hamiltonian $\hat{H}_\tf{el}$ includes both the electronic degrees of freedom and its coupling to the mechanical 
ones through
\begin{equation}
\hat{H}_\tf{el}(\hat{\bm{X}}) = \sum_\alpha E_\alpha(\hat{\bm{X}}) \ket{\alpha} \bra{\alpha}, 
\label{eq:Hel}
\end{equation}
where the sum runs over all possible electronic many-body $\alpha$-eigenstates.
In App.~\ref{app:1}, we show that any explicit $\bm{X}$-dependence on 
$\ket{\alpha}$ can be disregarded on the level of approximation we are going to take throughout this work.
\begin{figure}[h]
	\centering
	\includegraphics[width=\columnwidth]{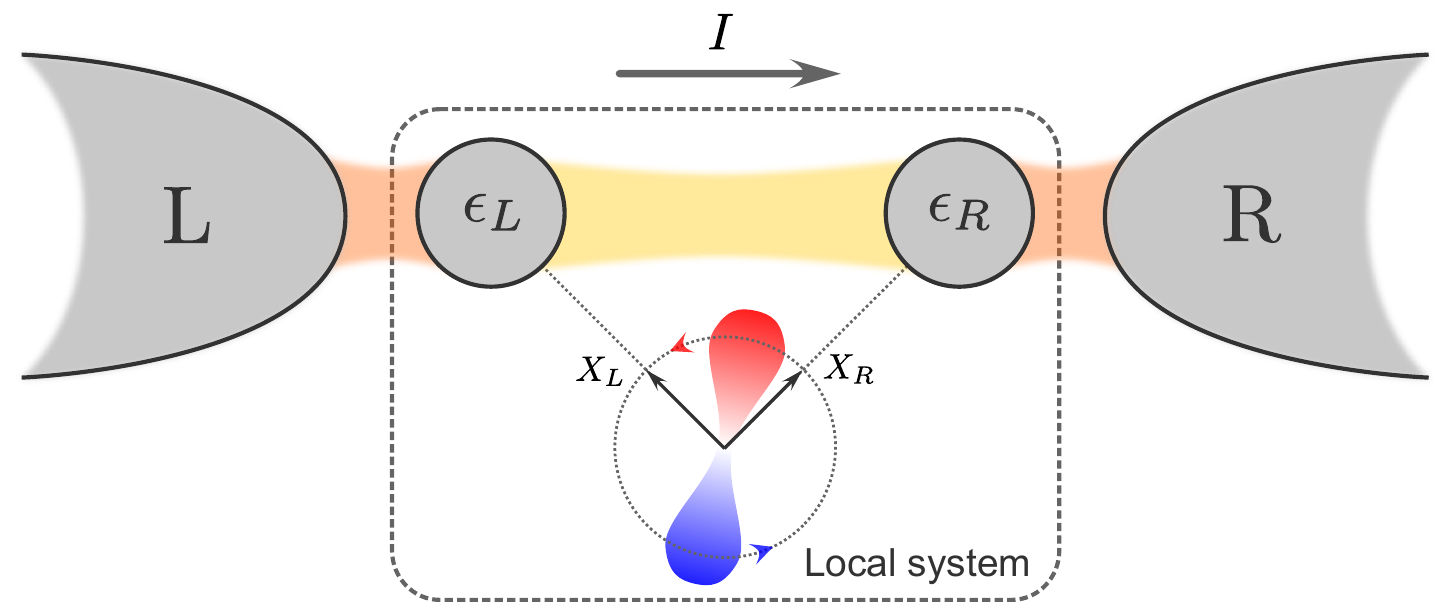}
	\caption{(Color online) Example of the type of system considered.	Here we show a double quantum dot capacitively coupled to an ideal rotor with 	fixed positive and negative charges. The current induced by a bias voltage leads to a force which may produce a rotational motion of the mechanical system (see Sec.~\ref{sec:dqd}). The local system (delimited by dashed lines) is assumed to be weakly coupled to the left (L) and right (R) reservoirs.}
	\label{fig:fig1}
\end{figure}
As schematically shown in Fig.~\ref{fig:fig1}, the local system is weakly coupled to left ($L$) 
and right ($R$) leads and the full Hamiltonian reads $\hat{H} = \hat{H}_\tf{sys} + \hat{H}_\tf{res} + 
\hat{H}_\tf{tun}$. The leads are described as reservoirs of noninteracting electrons through the Hamiltonian
\begin{equation}
\hat{H}_\tf{res} = \sum_r \hat{H}_r = \sum_{rk\sigma} \epsilon_{rk} 
\hat{c}_{rk\sigma}^\dag \hat{c}_{rk\sigma},
\end{equation}
where $\hat{c}_{rk\sigma}^\dag$ ($\hat{c}_{rk\sigma}$) creates (annihilates) an electron in 
the lead $r = L,R$ with spin $\sigma = \uparrow,\downarrow$ and state index $k$. As usual, the reservoirs are 
assumed to be at thermal equilibrium, characterized by a temperature $T$ and electrochemical potentials 
$\mu_L = -\mu_R = V/2$.~\footnote{Throughout this work we use $e=1$ for the absolute value of the electron 
charge and $\hbar = 1$.} The tunnel coupling between the local system and the leads is determined by the tunnel Hamiltonian
\begin{equation}
\hat{H}_\tf{tun} = \sum_{rk\sigma\ell} \left(t_{r\ell} \hat{d}_{\ell\sigma}^\dag \hat{c}_{rk\sigma} + 
\tf{H.c.} \right).
\end{equation}
Here $t_{r\ell}$ are the tunnel amplitudes which, for simplicity, we assume to be $k$ and $\sigma$ 
independent. The fermion operator $\hat{d}_{\ell\sigma}^\dag$ ($\hat{d}_{\ell\sigma}$) creates (annihilates) one 
electron in the single-particle state $\ell$ of the local system with spin $\sigma$. The tunnel-coupling strengths 
$\Gamma_{r\ell} = 2\pi \nu_r |t_{r\ell}|^2$ characterize the rate at which the tunnel processes take 
place. Here $\nu_r$ is the density of states in the $r$-lead, which is assumed to be energy-independent and 
with a band cutoff $D$, the largest energy scale. Note that $\hat{H}_\tf{el}$ was defined in the eigenstate 
basis while $\hat{H}_\tf{tun}$ is referred to the single-particle energy levels. Thus, the tunnel matrix 
elements accounting for transitions between different eigenstates are obtained as linear superpositions of the 
above tunnel amplitudes.~\cite{leijnse2008}
 
\textit{Langevin dynamics -} As a first step in the derivation of the dynamics of the mechanical system we 
start from the Heisenberg equation of motion for $\hat{\bm{P}}$,
\begin{equation}
m\ddot{\hat{\bm{X}}} + \hat{\nabla U}(\hat{\bm{X}},t) = -\hat{\nabla H}_\tf{el}(\hat{\bm{X}}). \label{eq:lang_op}
\end{equation}
The measured value of the involved operators can be taken as its mean value plus some fluctuation around it, 
i.e. $A = \braket{\hat{A}} + \xi_A$. We will work under the nonequilibrium Born-Oppenheimer approximation
~\cite{bennett2010,bode2011,thomas2012,bustos2013,holzbecher2014,arrachea2015,fernandez2015,ludovico2016,fernandez2017}
(or Ehrenfest approximation
~\cite{diventra2000,diventra2004,horsfield2004,dundas2009,todorov2010})
where the dynamics of the electronic and mechanical degrees of freedom are well separated and the latter can be treated classically.
This allows us to neglect fluctuations in the left hand side of Eq.~(\ref{eq:lang_op}) and then to obtain the following Langevin 
equation for the mechanical degrees of freedom:
\begin{equation}
m\ddot{\bm{X}} + \nabla U + \bm{F}_\tf{load} = \braket{\hat{\bm{F}}} + \bm{\xi}, \label{eq:lang}
\end{equation}
where $\braket{\hat{\bm{F}}} = - \braket{\hat{\nabla H}_\tf{el}} = i \braket{[\hat{H}_\tf{el}(\hat{\bm{X}}),\hat{\bm{P}}]}$ and $\bm{\xi}$ account for the mean value and the fluctuation of the CIF, respectively. Notice we have split the external
force [the force arising from the external potential in Eq.~(\ref{eq:Hsys})] into conservative ($-\nabla U$) and nonconservative ($\bm{F}_\tf{load}$) terms. This last plays the role of an eventual loading force, typically opposed to the mechanical motion (for this reason we use a minus sign in 
$\bm{F}_\tf{load}$). The main task therefore relies on the calculation of the expectation value of the 
CIF, which will be derived in Sec.~\ref{sec:method}. Once this force is obtained, we can use 
Eq.~(\ref{eq:lang}) to integrate the classical equations of motion and obtain $\bm{X}(t)$.
Finally, notice we are describing the motion of the mechanical degrees of freedom only through the mean value of $\bm{X}$, which is reasonable for large or massive objects. For smaller mechanical systems such as molecules or ions, however, some form of semiclassical approximation may be 
needed, see e.g. Ref.~\onlinecite{horsfield2004}.

\textit{Observables - } 
The time evolution of the expectation value of an arbitrary operator $\hat{R}$ is formally obtained by
\begin{equation}
R(t) = \braket{\hat{R}(t)} = \tr{} \, \hat{R} \, \hat{\rho}(t),
\label{eq:observable}
\end{equation}
where $\hat{\rho}(t)$ is the full system's density operator and the trace involves all electronic degrees of freedom. In this work, we focus on two observables: 
The charge \textit{tunnel} current $I_r(t) = 
\braket{\hat{I}_r(t)}$ entering the $r$-lead, and the current-induced force $\bm{F}(t) = 
\braket{\hat{\bm{F}}(t)}$ exerted on the mechanical degrees of freedom. Since in the decoupled system ($\hat{H}_\tf{sys}+\hat{H}_\tf{res}$) the number of particles 
is conserved, the operator related to the charge current is given by
\begin{equation}
\hat{I}_r = i[\hat{H}_\tf{tun},\hat{N}_r],
\end{equation}
where $\hat{N}_r$ is the number operator for the electrons in the reservoir $r$ and we use the sign convention 
that the particle current is positive when it flows towards the local system.
On the force side, the local coupling to the mechanical degrees of freedom enters through the eigenenergies of $\hat{H}_\tf{el}$.
Thus, the CIF only involves fermionic operators of the local system, such that its related observable can be obtained by tracing out the system's degrees of freedom:
\begin{equation}
\bm{F}(t) = \tr{sys} \, \hat{\bm{F}} \, \hat{p}(t), \label{eq:force}
\end{equation}
where $\hat{p}(t) = \tf{tr}_\tf{res} \, \hat{\rho}(t)$ is the reduced density operator of the local system.
In the next section, we will use the real-time diagrammatic approach of Ref.~\onlinecite{splettstoesser2006} to calculate $\hat{p}(t)$ and 
derive the explicit expressions for both the (local) force and the tunnel current expectation values.

\section{Real-time diagrammatic approach}
\label{sec:method}

The relevant part of the system's reduced density matrix, namely, its diagonal elements, can be obtained 
after tracing out the degrees of freedom of the leads. The time evolution of the occupation probabilities, 
represented by the vector $\bm{p}(t)$, is governed by the generalized master equation~\cite{splettstoesser2006}
\begin{equation}
 \frac{d}{dt}\bm{p}(t) = \int_{-\infty}^t dt'\bm{W}(t,t')\bm{p}(t'). \label{eq:kineq}
\end{equation}
The change in the occupation probabilities, due to electron tunnel processes between the local system 
and the leads, is described by the evolution kernel $\bm{W}(t,t')$. This kernel collects all irreducible 
diagrams in the Keldysh double contour~\cite{koenig1996} and its matrix elements $W_{\alpha \beta}(t,t')$ 
describe the transition from a state $\ket{\beta}$ at time $t'$ to a state $\ket{\alpha}$ at time $t$.
At the level of approximation we work here, the transport properties are completely determined by the 
diagonal elements [see Eq.~(\ref{eq:Hel})] of the reduced density operator.
The off-diagonal elements, related to coherent superpositions of different eigenstates, are decoupled from 
the diagonal ones due to charge and spin conservation in the tunnel event and/or a marked difference in 
their dynamical time scales. Therefore, they do not affect the observables of interest (i.e. charge tunnel 
current and CIF).

A charge current flow, due to a possibly fixed bias voltage, induces a periodic motion of the mechanical 
system. In particular, we will focus on systems where the mechanical freedom is able to reach a 
stationary regime characterized by a cyclic motion with period $\tau = 2\pi/\Omega$.
This mechanical motion, in turn, produces a modulation in the system's energies which leads to an additional pumping current. In this sense, the adiabatic expansion used in Refs.~\onlinecite{splettstoesser2006,riwar2010,juergens2013} to 
describe the pumping mechanism can be equally used here.
To this end, we will work in the adiabatic regime where the period $\tau$ of the mechanical modulation is larger than the typical time spent by the electrons inside the local system.
Strictly speaking, the frequency $\Omega$ and energy amplitude $\delta \epsilon$ associated with the mechanical motion are limited by the \textit{adiabaticity condition} $\Omega / \Gamma \ll k_\tf{B} T / \delta \epsilon$.
It is important to note that, unlike typical adiabatic pumping schemes where the modulation 
frequency can be controlled at leasure, the frequency of the mechanical motion is not well established from 
the ground up, and it depends on the system's parameters. Therefore, one should be careful in defining 
appropriate regimes where such a condition is fulfilled. When this is the case, the occupation probabilities can 
be expanded in powers of $\Omega$ as $\bm{p}(t) = \bm{p}^{(i)}+\bm{p}^{(a)}$. The first term (zeroth-order in $\Omega$) represents 
the {\em instantaneous} occupations and describes the steady state solution when the mechanical coordinates are frozen at time $t$. 
The instantaneous occupations are obtained from the time-dependent kinetic equation in the steady-state 
limit,~\footnote{This limit for the electronic time scale should not be confused with the above discussed 
mechanical stationary regime, the latter characterized by much longer times than those of the electronic 
degrees of freedom.} which to linear order in $\Gamma$ reads
\begin{equation}
\bm{0} = \bm{W} \bm{p}^{(i)}. \label{eq:kineqin}
\end{equation}
In this equation we introduced the zero-frequency Laplace transform of the instantaneous kernel $\bm{W} = 
\int_{-\infty}^t dt' \bm{W}^{(i)}(t-t')$.~\footnote{For the lowest order approximation in $\Gamma$ we take here, all kernels are instantaneous, and we therefore omit the $i$-superscript to simplify the notation.} 
The delayed response of the electronic degrees of freedom against the mechanical motion is collected by the next-to-leading term (linear in 
$\Omega$), $\bm{p}^{(a)}$, and obeys the following \textit{adiabatic} correction
\begin{equation}
 \frac{d}{dt} \bm{p}^{(i)} = \bm{W} \bm{p}^{(a)}. \label{eq:kineqad} 
\end{equation}
The occupation probabilities are then obtained by solving Eqs.~(\ref{eq:kineqin}) and 
(\ref{eq:kineqad}) together with the normalization conditions $\bm{e}^\tf{T}\bm{p}^{(i)} = 1$ and 
$\bm{e}^\tf{T}\bm{p}^{(a)} = 0$. Here, $\bm{e} = (1,...,1)^\tf{T}$ is a representation of the local system's 
trace operator. From Eq.~(\ref{eq:kineqad}), the adiabatic corrections to the occupation 
probabilities can be written in terms of the instantaneous contributions by
\begin{equation}
\bm{p}^{(a)} = \tilde{\bm{W}}^{-1} \frac{d}{dt} \bm{p}^{(i)}, \label{eq:occ_ad}
\end{equation}
where the (invertible) matrix $\tilde{W}_{\alpha \beta} = W_{\alpha \beta} - W_{\alpha \alpha}$ includes the 
normalization condition $\bm{e}^\tf{T} \bm{p}^{(a)} = 0$. Since the evolution kernel is linear in $\Gamma$ while the instantaneous 
occupations are $\mathcal{O}(\Gamma^0)$, the leading order adiabatic occupations are $\mathcal{O}(\Gamma^{-1})$. This, however, does not yield to any divergence as we always assume $\Omega/\Gamma < 1$.~\cite{splettstoesser2006}

The observables in Eq.~(\ref{eq:observable}) need to be equally expanded in both the frequency $\Omega$ and 
the tunnel-coupling strength $\Gamma$. Their results are then split into instantaneous and adiabatic 
parts
\begin{equation}
R^{(i/a)} = \braket{\hat{R}}^{(i/a)} = \bm{e}^\tf{T} \bm{W}^R \bm{p}^{(i/a)}, \label{eq:obs}
\end{equation}
where $\bm{W}^R$ is the instantaneous kernel of the corresponding observable $R$. For the charge current this 
kernel is linear in $\Gamma$ and writes $W^{I_r}_{\alpha \beta} = -(n_\alpha - n_\beta) W^r_{\alpha \beta}$, 
with $n_\alpha$ the number of particles in state $\ket{\alpha}$
and $\bm{W}^r$ the $r$-lead evolution kernel 
such that $\bm{W} = \sum_r \bm{W}^r$. We describe $R^{(a)}$ by a scalar product with the time-derivative of 
the local system's occupations
\begin{equation}
R^{(a)} = \bm{e}^\tf{T} \bm{W}^R \tilde{\bm{W}}^{-1} \frac{d}{dt} \bm{p}^{(i)} = 
\sum_\alpha \varphi_\alpha^R \frac{d}{dt}p_\alpha^{(i)}, \label{eq:adcurrent}
\end{equation}
with the sum running over the system eigenstates. Applied to the charge current, this equation expresses the 
response to a time-dependent variation in the instantaneous occupations induced by the mechanical modulation. 
The response coefficient $\varphi_{\alpha}^{I_r}$ determines the ratio at which the current $I_r$ 
flows into the $r$-lead due to a variation in the occupation of the state $\alpha$.

\subsection{Current-induced forces in interacting systems}
\label{subsec:force}

Due to the local parameter assumption that yields Eq.~(\ref{eq:force}), the ``kernel'' matrix associated with the $\nu$-component of the force is zeroth order in $\Gamma$ and its diagonal block simply writes as [see Eq.~(\ref{eq:local_diag}) in App.~\ref{app:3}]
\begin{equation}
W^{F_\nu}_{\alpha \beta} = - \frac{\partial E_\alpha}{\partial X_\nu}\delta_{\alpha \beta} 
\equiv F_{\nu,\alpha} \delta_{\alpha \beta}.
\label{eq:forceW}
\end{equation}
While in other formalisms the distinction between local and nonlocal observables can be somewhat arbitrary (see, e.g., Ref.~\onlinecite{bode2011}), for the tunnel coupling perturbation theory we use here this becomes crucial. In particular, the kernels associated with nonlocal forces might be quite different from the local ones, and for example the simple form of Eq.~(\ref{eq:force}) is no longer valid. 

As we mentioned before, we consider a Born-Oppenheimer regime where the mechanical coordinate enters as a 
classical variable. The CIF can now be expanded in terms of the velocity of the mechanical
coordinates ($\dot{\bm{X}} \propto \Omega$) in the same manner as in Eq.~(\ref{eq:adcurrent}), provided the 
mechanical velocity fulfills the adiabaticity condition. Hence, for the $\nu$-component one obtains
\begin{equation}
\braket{\hat{F}_\nu} = F^{(i)}_{\nu}+F^{(a)}_{\nu} = F^{(i)}_{\nu} - \sum_{\nu'} \gamma_{\nu\nu'} \dot{X}_{\nu'},
\end{equation}
where 
\begin{equation}
F^{(i)}_\nu = \bm{e}^\tf{T} \bm{W}^{F_\nu} \bm{p}^{(i)}, \qquad \gamma_{\nu\nu'} = -\bm{e}^\tf{T} 
\bm{W}^{F_\nu} \frac{\partial \bm{p}^{(a)}}{\partial \dot{X}_{\nu'}}, \label{eq:forcegamma}
\end{equation}
represent the instantaneous contribution to the force and the scalar elements of the friction tensor 
$\bm{\gamma}$, respectively. To fully characterize the CIF, later on we give a general expression for the force 
fluctuation in terms of the force correlation function. Here we used the $(i)$ and $(a)$ superscripts to denote that such quantities 
are instantaneous or adiabatic in the frequency expansion, respectively. In this sense, the electronic delay against the 
mechanical motion, entering through $\bm{F}^{(a)}$, can be thought as a frictional force that dissipates the
amount of energy delivered by the bias current. Importantly, these simple forms for the force terms come from 
the assumption of a local parameter modulation, given that the mechanical degrees of freedom are only present in the 
local system. Other modulation schemes including, e.g., the tunnel barriers ($\hat{H}_\tf{tun}$) or the 
electrochemical potentials ($\hat{H}_\tf{res}$), would involve the calculation of more involved
force-related kernels which are beyond the scope of this work.

Performing a line integral on Eq.~(\ref{eq:lang}) over a closed trajectory in the mechanical parameters 
yields
\begin{equation}
\oint \left( m\ddot{\bm{X}} + \nabla U + \bm{F}_\tf{load} \right) \cdot d\bm{X} =
\oint \braket{\hat{\bm{F}}} \cdot d\bm{X},
\end{equation}
where we assume an average process over trajectories, such that only the mean values survive. Recalling that in the left hand side of the equation only $\bm{F}_\tf{load}$ is non-conservative, we obtain the following stationary limit relation
\begin{equation}
\mathcal{W}_\tf{load}= \sum_\nu \oint \left(F_\nu^{(i)} - \sum_{\nu'} \gamma_{\nu\nu'} \dot{X}_{\nu'} 
\right) dX_\nu = \mathcal{W}_F - \mathcal{E}_\tf{dis}, \label{eq:Wext}
\end{equation}
which implies that, after one driving cycle, the loading work that the motor can perform consists of the difference between the instantaneous, current-induced work ($\mathcal{W}_F$) and the dissipated energy per period ($\mathcal{E}_\tf{dis}$). 

\subsection{Onsager's reciprocity relations}
\label{subsec:onsager}

In addition to the adiabatic expansion taken on the above observables, we could also think of a linear regime for the bias voltage or, more generally, the electrochemical potentials. In this case we can expand both the current and the force up to linear order in $\mu_r$ around the equilibrium where all reservoirs' temperatures and electrochemical potentials are set at the same level, i.e. $T_r = T$ and $\mu_r = \mu$:
\begin{align}
I_r &= I_{r,\tf{eq}}^{(i)} + \sum_{r'} \left. \frac{\partial I_r^{(i)}}{\partial \mu_{r'}} \right|_\tf{eq} \delta \mu_{r'} + \sum_{\nu'} \frac{\partial I_{r,\tf{eq}}^{(a)}}{\partial \dot{X}_{\nu'}} \dot{X}_{\nu'}, \\ 
F_\nu &= F_{\nu,\tf{eq}}^{(i)} + \sum_{r'} \left. \frac{\partial F_\nu^{(i)}}{\partial \mu_{r'}} \right|_\tf{eq} \delta \mu_{r'} + \sum_{\nu'} \frac{\partial F_{\nu,\tf{eq}}^{(a)}}{\partial \dot{X}_{\nu'}} \dot{X}_{\nu'}, \label{eq:expF}
\end{align}
with $\delta \mu_r = \mu_r - \mu$ the deviation from the equilibrium. In this expansion, the equilibrium instantaneous 
currents $I_{r,\tf{eq}}^{(i)}$ are always zero, while the equilibrium instantaneous force $\bm{F}_\tf{eq}^{(i)}$ can be finite but conservative. Since the occupation $p_{\alpha,\tf{eq}}^{(i)}$ is given by the Boltzmann factor $\exp(-E_\alpha/k_\tf{B}T)/z$, with $z$ the local system's partition function, it is easy to see that 
\begin{equation}
\bm{F}^{(i)}_\tf{eq} = -\nabla \psi, \qquad \psi = -k_\tf{B}T \ln(z), \label{eq:helmholtz}
\end{equation}
where $\psi$ is the local system Helmholtz's free energy. 

\begin{figure}[h]
	\centering
	\includegraphics[width=\columnwidth]{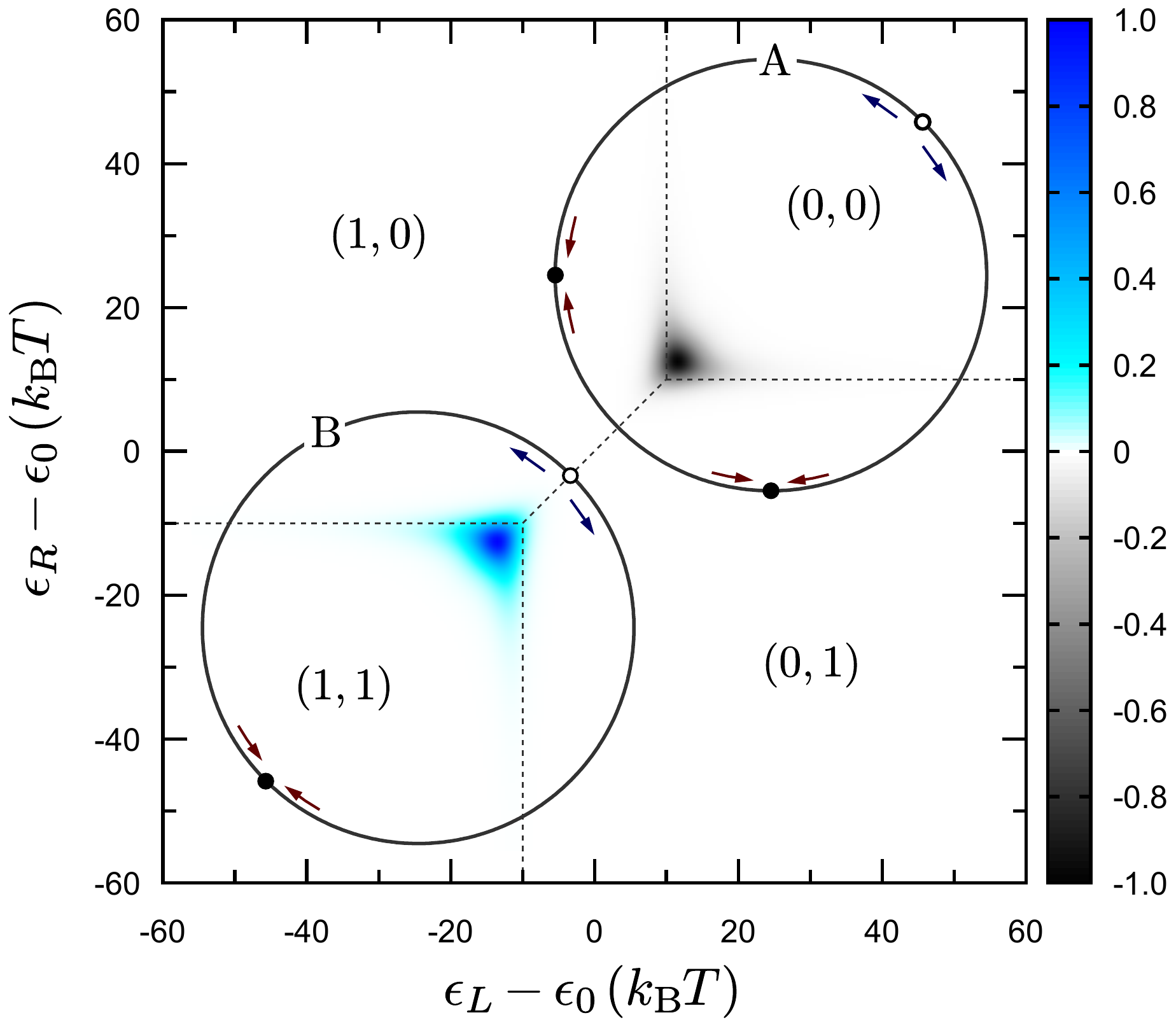}
  	\caption{(Color online) Normalized pseudomagnetic field $\mathcal{B}^I/\mathcal{B}^I_\tf{max}$ in energy domain [relative to the symmetry point 
   $\epsilon_0 = k_\tf{B}T \ln(2)-U/2$]. In Sec.~\ref{sec:dqd} we analyze two simple trajectories (marked as A and B in the figure) describing the motion of $\epsilon_L(\theta)$ and $\epsilon_R(\theta)$, where $\theta$ is the coordinate of the mechanical device [see Eq.~(\ref{eq:epsilontheta})]. The amount of pumped charge per cycle can be calculated through the surface integral of $\mathcal{B}^I$ along the area encircled by the trajectories. The charge regions $(n_L,n_R)$ are shown as reference in the $t_\tf{c} = 0$ limit and are delimited by dashed lines. The full (open) circles along the trajectories denote the minimum (maximum) potential associated with the rotor, see Eq.~(\ref{eq:Weff}). The used parameters are: $V = 0$, $U = 20~k_\tf{B}T$, $t_c = 5~k_\tf{B}T$, $\Gamma_L = \Gamma_R = \Gamma/2 = 0.25~k_\tf{B}T$, while for the trajectories we used: A) $\bar{\epsilon}_L = \bar{\epsilon}_R = -6~k_\tf{B}T + \delta \epsilon / \sqrt{2}$, B) $\bar{\epsilon}_L = \bar{\epsilon}_R = 2\epsilon_0+6~k_\tf{B}T - \delta \epsilon / \sqrt{2}$, with $\delta \epsilon = 30~k_\tf{B}T$.}
	\label{fig:fig2}
\end{figure}

In general terms, we can think of $-\mu_r$ and $\dot{X}_\nu$ as generalized forces ($x_i$) while $I_r$ and $F_\nu$ their associated fluxes 
($\phi_i$).~\footnote{The minus sign in $\mu_r$ comes from sign convention used for the charge current.} The above expansion thus writes:
\begin{equation}
\phi_i = \phi_{i,\tf{eq}}+\sum_j L_{ij}x_j, \qquad L_{ij}=\left. \frac{\partial \phi_i}{\partial x_j} \right|_\tf{eq},
\end{equation}
where the coefficients $L_{ij}$ are connected via Onsager's reciprocity relations, such that in the absence of magnetic fields they 
obey $L_{ij} = \pm L_{ji}$, and the sign depends on the adopted convention for the generalized forces and fluxes.~\cite{cohen2003,bustos2013,fernandez2015,ludovico2016,fernandez2017} We here prove that all these relations hold to lowest order 
in $\Gamma$ as far as the $L$-coefficients admit the following form
\begin{equation}
L_{ij} = \tf{constant} \times \sum_{\alpha\beta} W^{\phi_i}_{\alpha\beta} 
\left( \varphi^{\phi_j}_{\beta,\tf{eq}} - \bar{\varphi}^{\phi_j}_\tf{eq} \right) p^{(i)}_{\beta,\tf{eq}},
\label{eq:Lcoeff}
\end{equation} 
where $\bar{\varphi}_\tf{eq}^R = \bm{e}^\tf{T} \bm{W}^R \tilde{\bm{W}}^{-1} \bm{p}^{(i)}_\tf{eq}$ is the average $R$-response coefficient. If this 
is the case, as happens for $I_r$ and $F_\nu$, then we can use the following symmetry relation
\begin{equation}
\sum_{\alpha\beta}W^{\phi_i}_{\alpha\beta}(\varphi^{\phi_j}_{\beta,\tf{eq}}-\bar{\varphi}^{\phi_j}_\tf{eq})p^{(i)}_{\beta,\tf{eq}} = 
\sum_{\alpha\beta}W^{\phi_j}_{\alpha\beta}(\varphi^{\phi_i}_{\beta,\tf{eq}}-\bar{\varphi}^{\phi_i}_\tf{eq})p^{(i)}_{\beta,\tf{eq}}.  
\label{eq:ons_gral}
\end{equation}
As we show in App.~\ref{app:2}, this general relation relies on the \textit{detailed balance} property of the instantaneous occupations 
at equilibrium: $W_{\alpha\beta} p_{\beta,\tf{eq}}^{(i)} = W_{\beta\alpha} p_{\alpha,\tf{eq}}^{(i)}$.
In addition to Eq.~(\ref{eq:ons_gral}), we notice the following two important identities for the occupation derivatives in terms 
of the current and force response coefficients:
\begin{align}
\left. \frac{\partial p_\alpha^{(i)}}{\partial (-\mu_r)} \right|_\tf{eq} &= \frac{1}{k_\tf{B}T} 
\left( \varphi_{\alpha,\tf{eq}}^{I_r} - \bar{\varphi}^{I_r}_\tf{eq} \right) p_{\alpha,\tf{eq}}^{(i)}, \label{eq:occ_der1} \\
\left. \frac{\partial p_{\alpha}^{(a)}}{\partial \dot{X}_\nu} \right|_\tf{eq} &=
\frac{1}{k_\tf{B}T} \left( \varphi_{\alpha,\tf{eq}}^{F_\nu} - \bar{\varphi}^{F_\nu}_\tf{eq} \right) p_{\alpha,\tf{eq}}^{(i)}. \label{eq:occ_der2}
\end{align}
With these relations in mind, we now proceed with the crossed terms in the instantaneous current:
\begin{equation}
\left. \frac{\partial I_r^{(i)}}{\partial (-\mu_{r'})} \right|_\tf{eq} = \left. \frac{\partial I_{r'}^{(i)}}{\partial (-\mu_r)} \right|_\tf{eq},
\label{eq:ons_II}
\end{equation}
where obviously $r \neq r'$, otherwise the identity becomes trivial. For a symmetric bias this equation yields $(\partial (I_L^{(i)}+I_R^{(i)})/ \partial V)_\tf{eq} = 0$, in agreement with the instantaneous charge continuity equation.~\cite{juergens2013} By replacing Eq.~(\ref{eq:obs}) for the instantaneous current and noticing that $\bm{W}^r$ is independent of $\mu_{r'}$, the left term above writes
\begin{equation}
\sum_{\alpha\beta}W^{I_r}_{\alpha\beta} \left. \frac{\partial p^{(i)}_\beta}{\partial (-\mu_{r'})} \right|_\tf{eq} = \frac{1}{k_\tf{B}T}
\sum_{\alpha\beta}W^{I_r}_{\alpha\beta} \left( \varphi_{\beta,\tf{eq}}^{I_{r'}} - \bar{\varphi}^{I_{r'}}_\tf{eq} \right) p_{\beta,\tf{eq}}^{(i)},
\end{equation}
where we have used Eq.~(\ref{eq:occ_der1}) for the occupation derivative.
Now, from the general relation of 
Eq.~(\ref{eq:ons_gral}), we can interchange the observables, i.e. $I_r \leftrightarrow I_{r'}$ and arrive to 
the right hand side of Eq.~(\ref{eq:ons_II}). Continuing with the crossed terms, the adiabatic charge current 
should be related to the instantaneous components of the force via the following reciprocity relations:
\begin{equation}
\left. \frac{\partial F_\nu^{(i)}}{\partial (-\mu_r)} \right|_\tf{eq} = \frac{\partial I_{r,\tf{eq}}^{(a)}}{\partial \dot{X}_\nu}.
\label{eq:ons_FI}
\end{equation}
The force term in the left hand side can be easily written through its definition given in Eq.~(\ref{eq:obs})
\begin{equation}
\left. \frac{\partial F_\nu^{(i)}}{\partial (-\mu_r)} \right|_\tf{eq} 
= \frac{1}{k_\tf{B}T} \sum_{\alpha\beta} W^{F_\nu}_{\alpha\beta} \left( \varphi_{\beta,\tf{eq}}^{I_r} - \bar{\varphi}^{I_r}_\tf{eq} \right) p_{\beta,\tf{eq}}^{(i)},
\end{equation}
where we used the fact that the force kernel is a local system operator [see Eq.~(\ref{eq:forceW})], thus 
independent of $\mu_r$ and, as before, Eq.~(\ref{eq:occ_der1}) for the occupation derivative. Again, 
we can use Eq.~(\ref{eq:ons_gral}) to interchange the observables, i.e. $F_\nu \leftrightarrow I_r$, and 
through Eq.~(\ref{eq:occ_der2}) we arrive to the right hand side of Eq.~(\ref{eq:ons_FI}). Interestingly, 
we have obtained an equilibrium relation between terms coming from different orders in the frequency expansion. In the context 
of adiabatic pumping, it could be sometimes useful to keep in mind such a relationship to calculate the 
adiabatic pumped flux in terms of an instantaneous object. Performing a line integral of the 
instantaneous CIF over a closed trajectory $\partial \Sigma$, one obtains the useful work 
delivered by the bias current, i.e.
\begin{equation}
\mathcal{W}_F = \oint_{\partial \Sigma} \bm{F}^{(i)} \cdot d\bm{X} = \iint_\Sigma \nabla \times \bm{F}^{(i)} \cdot d\bm{S},
\end{equation}
where in the last equation we used Stokes' theorem. The work done by the bias current can then be represented 
either as the line integral of a pseudovector potential $\bm{\mathcal{A}}^F = \bm{F}^{(i)}$ or, alternatively, 
as the surface integral of a pseudomagnetic field $\bm{\mathcal{B}}^F = \nabla \times \bm{F}^{(i)}$. 
Such a representation of integral quantities in terms of auxiliary vector fields was also used in the context of adiabatic pumping and exploits here the geometric character of adiabatic quantum motors.
In Refs.~\onlinecite{calvo2012,yuge2012,haupt2013,juergens2013,riwar2013}, these vector 
fields were mathematical constructions from the line integral over the parameter trajectory defining pumped 
currents like charge, spin, heat, etc.
In particular, the charge pumped after one driving cycle can be written as the line integral of $\bm{\mathcal{A}}^{I_r} = \partial I^{(a)}_r/ \partial \dot{\bm{X}}$ or, alternatively, as the surface integral of a pseudomagnetic field $\bm{\mathcal{B}}^I = \nabla \times 
\partial I^{(a)}_r/ \partial \dot{\bm{X}}$.
In the linear bias regime, we can relate the force and the charge current vector fields through the above Onsager's reciprocity relation. From the $\mu_r$-expansion of Eq.~(\ref{eq:expF}) in the instantaneous force, we notice that its related pseudovector potential can be 
written as:
\begin{equation}
\bm{\mathcal{A}}^F = -\nabla \psi - \sum_r \bm{\mathcal{A}}^{I_r}_\tf{eq} \delta \mu_r,
\end{equation}
where we used Eq.~(\ref{eq:ons_FI}) and the fact that the equilibrium force is the gradient of the Helmholtz's free energy. Note that the pseudovector potential $\bm{\mathcal{A}}^{I_r}_\tf{eq}$ can be interpreted as the charge emissivity.~\cite{calvo2012,fernandez2017} Taking the curl at both sides we can relate the nonconservative part of the CIF to the pseudomagnetic field associated with the pumped charge, i.e.
\begin{equation}
\bm{\mathcal{B}}^F = -\sum_r \bm{\mathcal{B}}^{I_r}_\tf{eq} \delta \mu_r,
\label{eq:pseudoF}
\end{equation}
such that when integrated over the surface $\Sigma$ enclosed by the trajectory defined by the mechanical 
coordinates one arrives to
\begin{equation}
\mathcal{W}_F = -\sum_r Q_{I_r,\tf{eq}}^{(a)} \delta \mu_r = -Q_{I,\tf{eq}}^{(a)} V,
\label{eq:work_bias}
\end{equation}
where in the last term we defined $I = (I_L-I_R)/2$ due to the symmetric choice $\mu_L = -\mu_R = V/2$ and 
that no net charge is accumulated in the system after one period, i.e. $\sum_r Q_{I_r,\tf{eq}}^{(a)} = 0$. 
This simple relation between the work performed by the quantum motor and the pumped charge, already found 
in noninteracting systems described through the scattering matrix approach,~\cite{bustos2013,fernandez2015} 
also holds in systems with strong Coulomb interaction and weakly coupled to the leads. The obvious reason is that 
these two quantities are connected via the Onsager reciprocity relation of Eq.~(\ref{eq:ons_FI}). In 
Fig.~\ref{fig:fig2} we show the charge current pseudomagnetic field $\mathcal{B}^I_\tf{eq}$ for the 
double quantum dot system we discuss in Sec.~\ref{sec:dqd}. As stated by Eqs.~(\ref{eq:pseudoF}) and 
(\ref{eq:work_bias}), in the linear bias regime the trajectories that yield a nonzero $\mathcal{W}_F$ are 
those which enclose finite values of $\mathcal{B}^{I}_\tf{eq}$. In the figure these regions are close to 
the \textit{triple points} where three charge states are degenerate.~\cite{juergens2013} This motivates our 
later choice for the trajectories of the mechanical device such that the amount of work is maximized.
Obviously, the above relations between the force and charge current vector fields, together with 
$\mathcal{W}_F$ and $Q^{(a)}_{I,\tf{eq}}$, hold in the linear bias regime ($V \lesssim k_\tf{B}T$). 
For larger bias voltages, although such relations are no longer valid, one can still calculate 
all these quantities from the general definition given in Eq.~(\ref{eq:obs}).

The remaining reciprocity relations are
\begin{equation}
\frac{\partial F_{\nu,\tf{eq}}^{(a)}}{\partial \dot{X}_{\nu'}} = \frac{\partial F_{\nu',\tf{eq}}^{(a)}}{\partial \dot{X}_\nu},
\label{eq:ons_FF}
\end{equation}
and to prove them we can use Eqs.~(\ref{eq:occ_der2}) and (\ref{eq:ons_gral}) in the same way we proceeded 
before. Importantly, these relations imply the symmetric property of the friction tensor when it is evaluated in equilibrium, i.e. $\gamma_{\nu\nu'} = \gamma_{\nu'\nu}$.

\subsection{Fluctuation-dissipation theorem}
\label{subsec:FDT}

To complete the analysis of the force properties in equilibrium, we now derive the fluctuation-dissipation theorem between the force correlation function and the dissipation coefficients. In order to evaluate the force 
correlation we proceed in the same way as it was done in Refs.~\onlinecite{thielmann2003,riwar2013} for the 
zero-frequency current noise. The time-dependent force correlation $D_{\nu\nu'}(t)$ (or zero-frequency force noise), in our case, is defined as the 
time-integral of the two-time correlation function $D_{\nu\nu'}(t,t')$ by
\begin{equation}
D_{\nu\nu'}(t) = \int_{-\infty}^{\infty} dt' D_{\nu\nu'}(t,t'),
\end{equation}
where $D_{\nu\nu'}(t,t')=\braket{ \{ \hat{\xi}_\nu(t) , \hat{\xi}_{\nu'}(t') \} }$ and $\{ \bullet , \bullet \}$ denotes 
anticommutation. The force fluctuation operators $\hat{\xi}_\nu(t) = \hat{F}_\nu(t) - \langle \hat{F}_\nu(t) 
\rangle$ are written in the Heisenberg representation. 
In analogy with the zero-frequency current noise,~\cite{thielmann2003,riwar2013} this expression can be 
expanded in terms of $\Gamma$ and $\Omega$. Since in Eq.~(\ref{eq:lang}) we are considering the instantaneous 
fluctuations to lowest order in $\Gamma$, we show in App.~\ref{app:3} that the corresponding correlation term can be written as
\begin{align}
D_{\nu\nu'}^{(i)} &= \bm{e}^\tf{T} \bm{W}^{F_{\nu }} \bar{\boldsymbol{\Pi}} \bm{W}^{F_{\nu'}} \bm{p}^{(i)} + 
							\bm{e}^\tf{T} \bm{W}^{F_{\nu'}} \bar{\boldsymbol{\Pi}} \bm{W}^{F_{\nu }} \bm{p}^{(i)}, \label{eq:correlation} \\
\bar{\boldsymbol{\Pi}} &= \tilde{\bm{W}}^{-1} \left( \bm{p}^{(i)} \otimes \bm{e}^\tf{T} - \bm{1} \right). \nonumber
\end{align}
In our case where the force is a local system operator, their associated kernels are zeroth-order in $\Gamma$, 
while $\bar{\boldsymbol{\Pi}}$ is of order $\Gamma^{-1}$.~\cite{thielmann2003} In the time domain, this inverse dependence on the tunnel coupling strength indicates that the local correlations persist for longer times as the coupling to the leads goes to zero. Since $\Gamma$ is a perturbation parameter, one might think that the force fluctuations $\bm{\xi}$, related to these correlations, would diverge in this limit. However, as we discuss around Eq.~(\ref{eq:fluct}), the $\Gamma^{-1}$ dependence is compensated by a term $\Delta t$ accounting for the time-step of the numerical simulation.

As discussed above, the friction tensor $\bm{\gamma}$ in Eq.~(\ref{eq:forcegamma}) is related to the adiabatic contribution 
to the force. Their elements can also be computed as
\begin{equation}
\gamma_{\nu\nu'} = -\bm{e}^\tf{T} \bm{W}^{F_\nu} \tilde{\bm{W}}^{-1} \frac{\partial 
\bm{p}_\tf{eq}^{(i)}}{\partial X_{\nu'}}, \label{eq:gamma2}
\end{equation}
where we use Eq.~(\ref{eq:occ_ad}) for the adiabatic occupations and write $\dot{\bm{p}}^{(i)} = 
\sum_\nu (\partial \bm{p}^{(i)}/\partial X_\nu) \, \dot{X}_\nu$. For the derivatives of the occupations we use that in equilibrium these are Boltzmann factors and hence we 
can write
\begin{equation}
\frac{\partial \bm{p}_\tf{eq}^{(i)}}{\partial X_{\nu}} = - \frac{1}{k_\tf{B}T} \left( \bm{p}_\tf{eq}^{(i)} 
\otimes \bm{e}^\tf{T} - \bm{1} \right) \bm{W}^{F_{\nu}} \bm{p}_\tf{eq}^{(i)}, \label{eq:dpdx}
\end{equation}
such that
\begin{equation}
\gamma_{\nu\nu'} = \frac{1}{k_\tf{B}T} \bm{e}^\tf{T} \bm{W}^{F_\nu} \bar{\boldsymbol{\Pi}} \bm{W}^{F_{\nu'}} \bm{p}_\tf{eq}^{(i)}.
\end{equation}
According to Eq.~(\ref{eq:ons_FF}), the friction tensor is symmetric in equilibrium, meaning that the above expression is invariant under 
exchange of $\nu$ and $\nu'$ components. This allows us to compare with Eq.~(\ref{eq:correlation}) and obtain
\begin{equation}
\bm{D} = 2 k_\tf{B}T \bm{\gamma}, \label{eq:FDT}
\end{equation}
which indeed corresponds to the fluctuation-dissipation theorem for the force in lowest order in tunneling.
  
\section{Adiabatic quantum motor based on a double quantum dot}
\label{sec:dqd}

In this section we apply the above general results to a concrete example: An adiabatic
quantum motor based on a double quantum dot (DQD) with strong Coulomb interaction. Such 
a device is described through the following Hamiltonian:
\begin{eqnarray}
\hat{H}_\tf{el} &=& \sum_\ell \epsilon_\ell \hat{n}_\ell + U \hat{n}_L \hat{n}_R + \frac{U'}{2} 
\sum_\ell \hat{n}_\ell (\hat{n}_\ell - 1)  \nonumber \\
&& - \frac{t_\tf{c}}{2} \sum_\sigma (d^\dag_{L\sigma}d_{R\sigma} + \tf{H.c.}),
\end{eqnarray}
where $\hat{n}_\ell = \sum_\sigma d^\dag_{\ell\sigma} d_{\ell\sigma}$ is the $\ell$-dot particle 
number operator, with $\ell = L,R$. Here, the coupling with the mechanical degrees of freedom enters through 
the local energies $\epsilon_\ell$ of the dots. For simplicity we assume a linear dependence $\epsilon_\ell = 
\bar{\epsilon}_\ell + \lambda_\ell X_\ell$, where $\lambda_\ell$ sets the strenght of the coupling
between both mechanical and electronic degrees of freedom. $U$ and $U'$ are, respectively, the interdot and 
intradot charging energies. To simplify this analysis, we take the limit $U' \rightarrow \infty$, which 
forbids double occupation in a single dot. The last term accounts for the coupling between the two dots, and 
its strength is given by the hopping amplitude $t_\tf{c}$. The eigenstates of this Hamiltonian can be 
obtained after diagonalization 
of the single-particle block, which yields bonding $\ket{\tf{b}\sigma} = \hat{d}^\dag_{\tf{b}\sigma}\ket{0}$ and 
antibonding $\ket{\tf{a}\sigma} = \hat{d}^\dag_{\tf{a}\sigma}\ket{0}$ states with eigenenergies
\begin{equation}
E_\tf{b/a} = \frac{\epsilon_L + \epsilon_R}{2} \mp \sqrt{ \left(\frac{\epsilon_L - 
\epsilon_R}{2}\right)^2 + \left(\frac{t_\tf{c}}{2}\right)^2}.
\end{equation}
Two important remarks need to be noticed in what follows: First, the double-dot eigenbasis actually \textit{depends} 
on the mechanical coordinates $X_\ell$. Second, coherent superpositions of $\ket{\tf{b}\sigma}$ and 
$\ket{\tf{a}\sigma}$ states entering through off-diagonal elements of $\hat{p}(t)$ could in principle play a 
role. We assume, however, a strong interdot coupling regime~\cite{wunsch2005,riwar2010,juergens2013} where 
$t_\tf{c} \gg \Gamma$, such that these two related effects can be disregarded to lowest order in the tunnel 
coupling. In App.~\ref{app:1} we discuss this in more detail.
The many-body eigenstates can thus be constructed by adding electrons in the bonding or antibonding states 
and the DQD reduced density matrix writes (in vector form) as $\bm{p} = (p_0, p_{\tf{b}\uparrow}, 
p_{\tf{b}\downarrow}, p_{\tf{a}\uparrow}, p_{\tf{a}\downarrow}, p_{\uparrow\uparrow}, p_{\uparrow\downarrow},
p_{\downarrow\uparrow}, p_{\downarrow\downarrow})^\tf{T}$. The vector components thus represent the probabilities for the 
DQD either empty ($p_0$), singly occupied with an electron with spin $\sigma$ in the bonding ($p_{\tf{b}\sigma}$) or 
antibonding state ($p_{\tf{a}\sigma}$) or doubly occupied ($p_{\sigma\sigma'}$), where $\sigma$ and $\sigma'$ label the spin of the electrons in the left and the right dots, respectively.

\subsection{Physical model and trajectory}

As illustrated in Fig.~\ref{fig:fig1}, a possible~\footnote{Another example would be that of a carbon 
nanotube based quantum dot~\cite{benyamini2014} where the mechanical parameter $X_\alpha$ measures the 
distance between the $\alpha$-dot to the gate contact. In this case, the nanotube needs to be coupled to two independent vibrational modes in order to have nonzero adiabatic pumped charge per cycle and useful work.} example for a mechanical device in this type of systems would be that of an ideal electric rotor: A dipolar configuration of electric charges, which can perform a rigid rotation around its center.
Given the proximity between the electronic and mechanical subsystems, an electron that flows through 
the DQD in response to a bias voltage gives part of its impulse to the rotor. Such impulse produces a 
rotation of the mechanical system which, in turn, modifies the energies of the dots as it would be done by local gates.
To describe the motion of the rotor, we can take as mechanical coordinate the angle $\theta$ describing its orientation. The exact dependence of the dots' eigenenergies on $\theta$ will be given by the precise positioning of the rotor with respect to the dots. For simplicity let us assume the following dependence:
\begin{align}
  \epsilon_L (\theta) &= \bar{\epsilon}_L + \delta \epsilon \cos \theta, \nonumber \\
  \epsilon_R (\theta) &= \bar{\epsilon}_R + \delta \epsilon \sin \theta. \label{eq:epsilontheta}
\end{align}
In the energy domain, the above equations define a circular trajectory of radius $\delta \epsilon$ centered at the working point ($\bar{\epsilon}_L, \bar{\epsilon}_R$) as shown in Fig. \ref{fig:fig2}.
The mean energies $\bar{\epsilon}_{\ell}$ can be thought independent of the mechanical coordinate and, therefore, able to be controlled by external gate voltages. Then, one can ask for a convenient choice for the working point and $\delta \epsilon$. In our case, we are interested in maximizing the amount of useful work delivered by the bias current. From Eq.~(\ref{eq:work_bias}) we know 
that, in the linear bias regime, this quantity increases with the amount of adiabatic pumped charge. 
Therefore, we can first calculate the pseudomagnetic field $\mathcal{B}^I$ associated with the adiabatic 
charge current to exploit its geometric form and, with it, maximize the amount of work in one cycle of the 
parameters' trajectory.

In Fig.~\ref{fig:fig2} we show the normalized $\mathcal{B}^I$ at zero bias together with the considered 
trajectories in the energy domain.
This field coincides with that calculated in Ref.~\onlinecite{juergens2013} and is only nonzero around the 
triple degeneracy points. We will focus on trajectory A, which involves transitions between the empty 
and single particle states; and trajectory B,  involving transitions between single and double particle states. 
In both cases, the amount of pumped charge per cycle is close to 
one electron charge in magnitude,~\cite{riwar2010} and its sign depends on the direction of rotation of 
the mechanical system. Interestingly, the sign difference in the peaks of $\mathcal{B}^I$ (also present in 
$\mathcal{B}^F$) implies that, for a fixed bias, the motor working in trajectory A rotates in the opposite direction
as it would do in trajectory B.

\subsection{Angular Langevin equation} 

To describe the dynamics of the system, we start by projecting the Langevin equation [Ec.~(\ref{eq:lang})] on the circular trajectory defined in the space of parameters. In this situation, the only relevant direction is the tangential one, given by the unit vector $\hat{\bm{\theta}}$, since all radial forces are assumed to be compensated each other. In other words, the rotor radius is assumed to be time-independent. Working with polar coordinates, we obtain an effective Langevin equation for the angular coordinate of the rotor in terms of rotational forces, i.e.
\begin{equation}
  \ddot{\theta} = \frac{1}{\mathcal{I}} [ \mathcal{F}^{(i)}_{\theta} -\frac{\partial U}{\partial \theta} - 
  \mathcal{F}_\tf{load} - \gamma_{\theta} \dot{\theta} + \xi_{\theta}],
  \label{eq:lang-theta}
\end{equation}
where $\mathcal{I}$ is the moment of inertia associated with the mechanical system, 
$\mathcal{F}_\tf{load} = \bm{F}_\tf{load} \cdot \hat{\bm{\theta}}$ and
\begin{equation}
	\mathcal{F}^{(i)}_{\theta} = - \sum_\alpha \frac{\partial E_\alpha}{\partial
	\theta} p^{(i)}_\alpha, \quad 
	\gamma_\theta = \sum_{\alpha \beta}
	\frac{\partial E_\alpha}{\partial \theta} \tilde{W}^{-1}_{\alpha \beta}
	\frac{\partial p^{(i)}_\beta}{\partial\theta},
\end{equation}
are the current-induced torque and its associated friction term, respectively. In general, the fluctuation 
terms in the force $\xi_{\nu}$ are obtained from the elements $D_{\nu \nu'}$ of the force correlation matrix. 
Since in this case we project on the tangential direction, we can deduce from Eq.~(\ref{eq:correlation}) the 
correlation in the torque in terms of the angular variable through
\begin{equation}
	\mathcal{D}_\theta = - \sum_{\alpha \beta} \frac{\partial E_\alpha}{\partial\theta} 
	\left[ \mathcal{F}^{(i)}_\theta + \frac{\partial E_\beta}{\partial\theta} \right] \tilde{W}^{-1}_{\alpha \beta} p^{(i)}_\beta, 
\end{equation}
and with this quantity we can obtain the fluctuation term $\xi_{\theta}$. Since this last will be represented as a stochastic variable, we will use along this work the following expression
\begin{equation}
	\xi_\theta(s) = g(s) \sqrt{\frac{\mathcal{D}_\theta}{\Delta t}},
\label{eq:fluct} 
\end{equation}
where $g(s)$ represents a random value extracted from a standard normal distribution. The term $\Delta t$ is the discrete time step employed in the evolution algorithm and it accounts for the fluctuation averaging process in time. The idea behind this parameter is the following: If we take 
$\Delta t$ small, then the stochastic processes cannot be averaged enough and the randomness in $\xi_{\theta}$ becomes large; if $\Delta t$ is large, between two steps of the algorithm these stochastic processes are self-averaged, yielding a small $\xi_{\theta}$ contribution. 
As for the numerical integration of Eq.~(\ref{eq:lang-theta}) we assume force correlations which are local in time, i.e. $D_\theta(t,t') \simeq D_\theta \delta(t-t')$, the time step needs to be larger than the typical relaxation of the local correlation function obtained in Eq.~(\ref{eq:correlation}). This implies that $\Delta t > 1/\Gamma$. The fact that $\Delta t$ enters in the squared root ensures that the influence of the fluctuation on $\theta$ and $\dot{\theta}$ becomes independent of the time step.

\begin{figure*}[t]
	\centering
	\includegraphics[width=\textwidth]{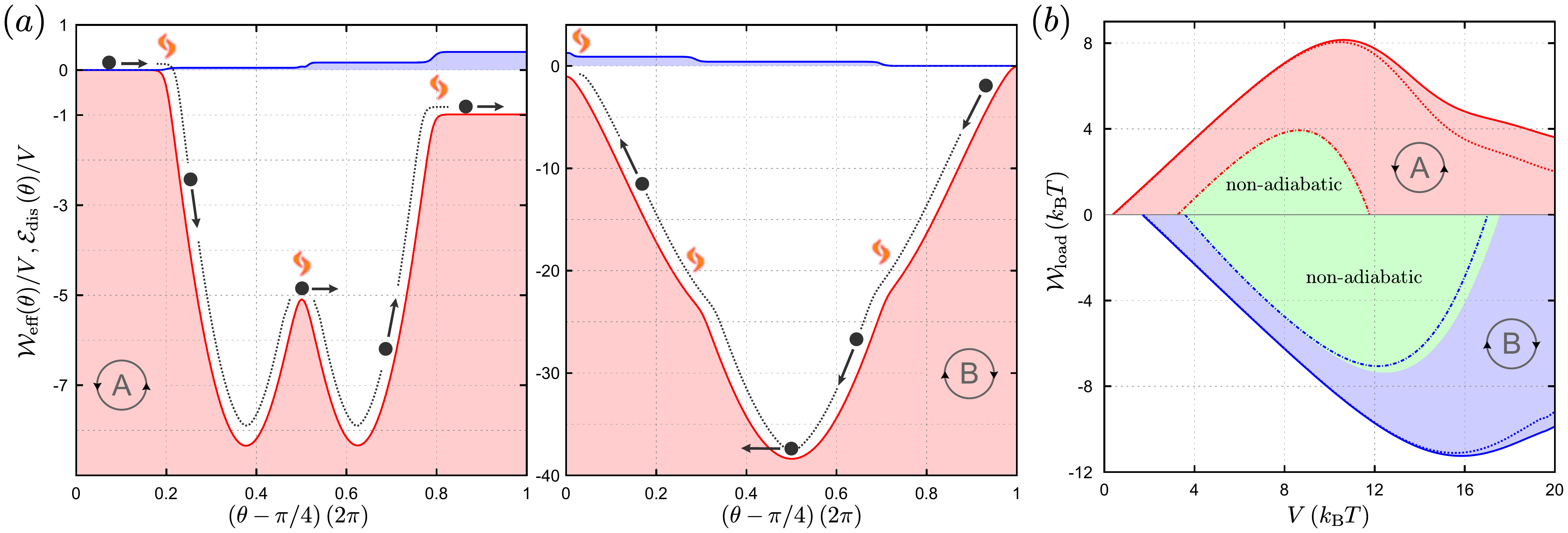}
	\caption{(Color online) (a) Effective work (red) and dissipated energy (blue) along one cycle of trajectories A (left) and B (right) in units of the used bias voltage $V = 2~k_\tf{B}T$ and for zero loading force. (b) Operation regimes of the motor as function of $\mathcal{W}_\tf{load}$ and $V$ for trajectories A ($\mathcal{W}_\tf{load} \geq 0$, red) and B ($\mathcal{W}_\tf{load} \leq 0$, blue). The shaded red and blue areas denote the regions where the motor works properly in the adiabatic regime. The crossover between these regions and the non-adiabatic ones (green areas) were calculated numerically in the time domain, and the dashed dotted lines follow the adiabaticity condition $\Omega/\Gamma < k_\tf{B}T/\delta \epsilon$, with $\Omega$ estimated from	Eq.~(\ref{eq:estimacion}). The dashed lines correspond to the first order estimation made by Eq.~(\ref{eq:condition1}), while the solid lines correspond to a higher-order estimation discussed in App.~\ref{app:4}. All other parameters coincide with those of Fig.~\ref{fig:fig2} and we used $\mathcal{I}= 750~k_\tf{B}T/\Gamma^2$.}
\label{fig:fig3}
\end{figure*}

To focus on the CIF part of Eq.~(\ref{eq:lang-theta}) and to give a simple description of the operation of the motor, we will neglect in what follows the role of the external conservative force $-\partial_\theta U$ as it does not contribute to the overall work per cycle and it depends on the detailed interaction of the rotor with its surroundings. Additionally, we want a simple expression for $\mathcal{W}_\tf{load}$ 
and therefore we limit to the case in which $\mathcal{W}_\tf{load}$ is constant and independent of $\dot\theta$. This can be associated with processes such as formation of chemical bonds as in the case of biological nanomotors.~\cite{goel2008,guix2014}
For simplicity we will consider only a constant loading force along the tangential direction, i.e. $\bm{F}_\tf{load} = F_\tf{load} \hat{\bm{\theta}}$. In this way, the associated loading work simply results $\mathcal{W}_\tf{load} = 2 \pi \mathcal{F}_\tf{load}$, where $\mathcal{F}_\tf{load}$ is indeed a torque as the rest of the forces in Eq.~(\ref{eq:lang-theta}). Other models for $\mathcal{W}_\tf{load}$ involving, for example, a mechanical dissipation, are also possible within this frame and, in such a case, might enter as a renormalization of the friction coefficient $\gamma_\theta$.

\subsection{Operational regime of the motor}
\label{subsec:regimes}

To gain some intuition on the dynamical behavior of the motor, in Fig.~\ref{fig:fig3}(a) we show the 
effective work
\begin{equation}
\mathcal{W}_\tf{eff}(\theta) = \int_0^\theta \left( \mathcal{F}_\tf{load} - \mathcal{F}^{(i)}_{\theta'} \right) d\theta',
\label{eq:Weff}
\end{equation}
together with an estimation of the amount of dissipated energy along one cycle of trajectories A (left) and B (right). Here we neglect force fluctuations to simplify the following qualitative analysis, though they will be later included in Sec.~\ref{subsec:dynamics} when describing the motor's dynamics. We use a negative sign in 
$\mathcal{W}_\tf{eff}(\theta)$ to mimick the above integral as a potential energy term. For trajectory A, this function renders a double well 
potential in $\theta$, with the wells located at $\theta = \pi$ and $3\pi/2$ [see full circles in Fig.~\ref{fig:fig2}(a)], respectively, and an internal barrier in $\theta = 5 \pi/4$, whose height indeed depends inversely on $t_c$. The shape of $\mathcal{W}_\tf{eff}$ suggests that if the rotor is initially located around the first plateau ($\theta \simeq \pi/4$) and it slowly rotates in the anticlockwise direction then, eventually, 
it will arrive to a depletion region where its angular velocity suddendly increases. This occurs when the DQD picks up an electron from the left lead. The gained kinetic energy then allows the rotor to cross the barrier between the two wells, meaning that the electron located in the left dot 
tunnels into the right dot. If the bias voltage is strong enough, then the rotor arrives to a second plateau ($\theta \simeq \pi/4 + 2\pi$) where the electron leaves the DQD towards the right lead. In this case, the rotor's final angular velocity is larger than the initial one. In fact, the energy difference $\Delta\mathcal{W}_\tf{eff}$ between two successive plateaux is proportional to the bias voltage [see Eq.~(\ref{eq:work_bias})] and yields the motion of the motor (represented by a black dot in the figure). With this simple analysis, we established, at least qualitatively, the connection between the work done by the motor and the amount of pumped electrons per cycle. Indeed, the latter only depends on the occupation sequence performed along the cycle. As for trajectory A and $V>0$ this is $(0,0) \rightarrow (1,0) \rightarrow (0,1) \rightarrow (0,0)$, the total number of pumped particles (on top of the instantaneous current) is one electron from left to right. For trajectory B and $V>0$, $\mathcal{W}_\tf{eff}$ shows a single well much deeper than those of trajectory A. This is attributed to the fact that there is always one or two electrons occupying the DQD during the cycle, and the CIF (in this model) is proportional to the occupation number in the local system. As in this case the motor rotates in the clockwise direction, the occupation sequence is $(0,1) \rightarrow (1,1) \rightarrow (1,0) \rightarrow (0,1)$, so again we obtain the same amount (and sign) of pumped particles per cycle.

The rotor also dissipates part of its energy at different points of the cycle, characterized by transitions between different charge regions $(n_L,n_R)$ (see dashed lines in Fig.~\ref{fig:fig2}). This means that the rotor can move freely within these regions and each time a tunnel event occurs, a certain amount of kinetic energy is lost through dissipation, as shown by the blue curves~\footnote{The dissipated energy in the figure 
was obtained through a first-order recursion formula for $\dot{\theta}$, see App.~\ref{app:4}.} in Fig.~\ref{fig:fig3}(a). Consequently, after a certain number of cycles the rotor arrives to a stationary regime where the difference in $\dot{\theta}$ between two successive plateaux becomes negligible. This regime, nevertheless, is not always guaranteed if the dissipation is strong enough as to prevent the rotor to reach the second plateau. In this case the rotor gets stuck in the depletion region and can no longer complete the cycle. When this occurs, the final trajectory in parameter space is just some arc of the full circle and no area is enclosed, such that the motor can no longer perform useful work.

To determine in which of these two regimes will the rotor end up, we start from Eq.~(\ref{eq:Wext}) where we 
related the work per cycle performed by the CIF with the amount of dissipated energy and a 
possible extra loading work. The total work is therefore
$\mathcal{W}_\tf{tot} = \mathcal{W}_F - \mathcal{E}_\tf{dis} - \mathcal{W}_\tf{load}$. Taking into account 
the above rotational forces, this can be expressed as
\begin{equation}
\mathcal{W}_\tf{tot} = \int_0^{2\pi} \left[ \mathcal{F}^{(i)}_\theta - \mathcal{F}_\tf{load} - 
\gamma_\theta \dot{\theta} \right] d\theta.
\label{eq:wtot}
\end{equation}
As we already mentioned, once the stationary regime is reached these quantities equate and yield
$\mathcal{W}_\tf{tot} = 0$. To arrive to this situation, however, this quantity needs to be always positive. 
This determines the operation condition of the motor, i.e. $\mathcal{W}_F - \mathcal{W}_\tf{load} \geq 
\mathcal{E}_\tf{dis}$, as the motor reaches the stationary regime. We notice that the dissipation term 
depends on the angular velocity $\dot{\theta}$ which, in principle, is 
not known. In App.~\ref{app:4} we derive a recursive formula to solve $\dot{\theta}$ as function of $\theta$. 
To first order in the recursion, this yields the following condition
\begin{equation}
	\mathcal{W}_F-\mathcal{W}_\tf{load}^{*} = \int_0^{2\pi} \gamma_{\theta}\sqrt{\frac{2}{\mathcal{I}} \int_0^\theta 
	\left( \mathcal{F}^{(i)}_{\theta'} - \mathcal{F}_\tf{load}^{*} \right) d\theta'} d\theta,
	\label{eq:condition1}
\end{equation}
where $\mathcal{W}_\tf{load}^{*} = 2\pi\mathcal{F}_\tf{load}^{*}$ is the maximum allowed loading work such 
that the motor can move indefinitely towards the stationary regime. So, for a given value of the bias 
voltage, we can calculate both $\mathcal{F}^{(i)}_\theta$ and $\gamma_\theta$ along one period and then use 
the above equation to obtain $\mathcal{W}_\tf{load}^{*}$ numerically. In Fig.~\ref{fig:fig3}(b) we show the 
allowed values of $V$ and $\mathcal{W}_\tf{load}$ for which the motor reaches the stationary regime in red and 
blue shaded regions for trajectories A and B, respectively. These regions were obtained by evaluating the operation 
condition through the numerical solution of Eq.~(\ref{eq:lang-theta}) in time domain. 
In dashed red (blue) we show the estimation given by Eq.~(\ref{eq:condition1}) for trajectory A (B), and is accurate up to 
$V \simeq 9~k_\tf{B}T$ ($V \simeq 15~k_\tf{B}T$). For larger bias values this line no longer fits the crossover and one needs 
to consider higher orders in the recursive solution, as the solid red (blue) curve corresponding to the fourth (fifth) order 
solution (see App.~\ref{app:4}).

Importantly, the adiabatic expansion discussed in Sec.~\ref{sec:method} needs to be consistent 
with the type of solution obtained from Eq.~(\ref{eq:lang-theta}). This implies that the adiabaticity condition $\Omega/\Gamma 
< k_\tf{B}T / \delta \epsilon$ needs to be fulfilled once the stationary regime is reached. In shaded green we 
show the nonadiabatic regions obtained from the numerical solution of Eq.~(\ref{eq:lang-theta}). This case can
be interpreted as follows: Just a small fraction of the amount of energy delivered by the bias current is 
dissipated per cycle and cannot prevent the rotor to move in a time scale which is comparable with that of 
the electrons flowing through the DQD. To have a simple test without recurring to the numerical 
time-evolution of $\dot{\theta}$, we can consider Eq.~(\ref{eq:wtot}) in the stationary regime where 
$\mathcal{W}_\tf{tot} = 0$ and take $\dot{\theta} = \Omega$ constant along the whole period. This is a rough 
approach since there is some obvious variation of $\dot{\theta}$ we are neglecting as the rotor completes one 
cycle, as suggests Fig.~\ref{fig:fig3}(a). Nevertheless, this approach is accurate enough for our purposes 
as we only want to compare the rate at which the rotor moves with $\Gamma$. Under this approach, we thus 
obtain
\begin{equation}
\Omega = \frac{\mathcal{W}_F - \mathcal{W}_\tf{load}}{2\pi\bar{\gamma}}, \qquad \bar{\gamma}=\int_0^{2\pi} 
\gamma_\theta \frac{d\theta}{2\pi}.
\label{eq:estimacion}
\end{equation}
In Fig.~\ref{fig:fig3}(b) we show such a crossing where $\Omega/\Gamma < k_\tf{B}T / \delta \epsilon$ is no longer 
fulfilled (dashed dotted lines), and hence the adiabatic expansion, up to first order in frequency, can no longer 
describe the motion of the rotor properly. As can be seen, in trajectory A this estimation fits very well with the 
numerical crossover between the two shaded regions, while in trajectory B some deviation appears in the large bias regime.

All in all, by increasing the bias voltage we ensure the operation of the device in the sense that the rotor 
reaches a stationary regime where it moves indefinitely. To ensure the validity of the adiabatic approximation, 
however, it may be necessary to ``slow down'' the rotor by including a loading force term.
Interestingly, in the large bias regime this is not always necessary, as we can see from Fig.~\ref{fig:fig3}(b), where 
for $V \gtrsim 12~k_\tf{B}T$ (trajectory A) and $V \gtrsim 18~k_\tf{B}T$ (trajectory B) the adiabatic condition is 
fulfilled even for $\mathcal{W}_\tf{load} \simeq 0$. We also observe in this regime that the maximum allowed 
loading work (solid line) decreases with $V$. This is due to deviations in the linear dependence of the current 
induced work with bias. In fact, $\mathcal{W}_F$ decreases with $V$ due to strong deformations of $\mathcal{B}^F$. In 
any case, as we discuss in the next section, the efficiency and the output power of the motor are strongly suppressed at 
large biases since almost all the work is lost through dissipation.

\subsection{Dynamics of the motor}
\label{subsec:dynamics}

In order to study the dynamics of the system we need to solve Eq.~(\ref{eq:lang-theta}). To this end, we 
set as starting point an initial position such that $\mathcal{W}_\tf{eff}$ 
is maximum (or, equivalently, $\mathcal{F}_\theta^{(i)} = \mathcal{F}_\tf{load}$) and 
then we consider a small initial velocity to slightly move the motor from the unstable equilibrium position. 
In each time step the values of $\mathcal{F}^{(i)}_\theta$, $\gamma_\theta$ and $\mathcal{D}_\theta$ may be 
obtained by interpolation to reduce the computing time. Once the variables $\theta(t)$ and $\dot{\theta}(t)$ 
are obtained, we proceed with the evaluation of other quantities like $\mathcal{W}_F$, $\mathcal{E}_\tf{dis}$, 
etc.
\begin{figure}[!h]
   \centering
   \includegraphics[width=.95\columnwidth]{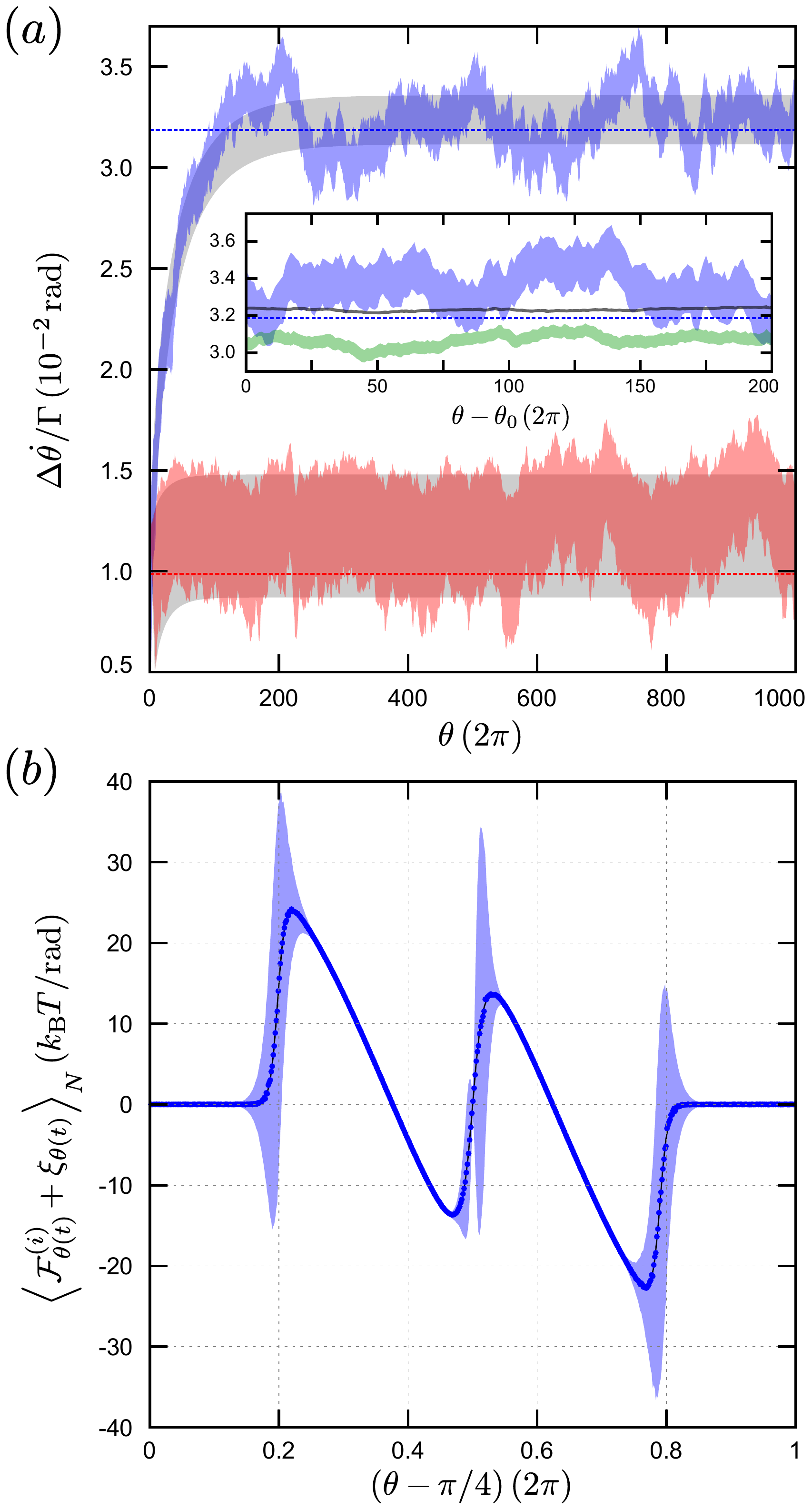}
\caption{(Color online) (a) Angular velocity range (taken as the minimum to maximum values of $\dot{\theta}$ over one period) divided by $\Gamma$ for the bias voltages and loading forces: $V = 2~k_\tf{B}T$ and 
$\mathcal{F}_\tf{load} = k_\tf{B}T / 2\pi$ (red) and $V = 8~k_\tf{B}T$ and 
$\mathcal{F}_\tf{load} = 4~k_\tf{B}T / 2\pi$ (blue). The reference ranges where the fluctuation is 
neglected are shown in gray in both cases. The estimated values of $\Omega$ are shown in 
dashed red and blue, respectively. Inset: Angular velocity ranges in the stationary regime for 
$V = 8~k_\tf{B}T$ and $\mathcal{F}_\tf{load} = 4~k_\tf{B}T / 2\pi$ and three different moments of 
inertia: $\mathcal{I} = \mathcal{I}_0$ (blue), $4 \mathcal{I}_0$ (green), and 
$16 \mathcal{I}_0$ (black), where $\mathcal{I}_0 = 750~k_\tf{B}T/\Gamma^2$ is the used 
moment of inertia in all other figures. To reach the stationary regime in each case we used 
$\theta_0/2\pi = 1000~\mathcal{I}/\mathcal{I}_0$. The estimated value $\Omega$ is shown in dashed 
blue.	(b) Averaged instantaneous, current-induced torque (blue dots) after $N = 4000$ realizations in the stationary regime. The solid line shows the reference case where the fluctuations are neglected, while the shaded region (blue) shows its standard deviation due to $\xi_\theta$. The chosen bias voltaje and loading force are $V = 2~k_\tf{B}T$ and $\mathcal{F}_\tf{load} = 0$, respectively, while the rest of the  parameters coincide with those of Fig.~\ref{fig:fig2} (trajectory A).}
   \label{fig:fig4}
\end{figure}
In Fig.~\ref{fig:fig4}(a) we show an example of the time evolution of the rotor's angular velocity for two different bias voltages in trajectory A. To avoid cluttering we show, in each 
cycle, the minimum and maximum values of $\dot{\theta}$, which allows us to visualize the internal 
range of velocities over time. These ranges are represented by shaded regions and we take as reference 
(in gray) the cases where the fluctuations are neglected. We can observe how the system reaches the 
stationary regime when these ranges become constant. The time spent for the rotor to 
arrive to this regime (stabilization time) is proportional to the moment of inertia $\mathcal{I}$, as 
suggested by Eq.~(\ref{eq:lang-theta}). Larger values of $\mathcal{I}$ imply a more pronounced separation between electronic and mechanical time-scales, which translates in a slower variation of $\dot{\theta}$ between two successive cycles. In consequence, when increasing $\mathcal{I}$ it takes to the rotor more time or, equivalently, a larger number of cycles to reach the stationary regime. 
Another effect of increasing $\mathcal{I}$ is that it reduces the velocity fluctuations of the rotor as it becomes evident from Eq.~(\ref{eq:lang-theta}). This is shown in the inset of Fig.~\ref{fig:fig4}(a) for three different values of $\mathcal{I}$.

As can be inferred from Eq.~(\ref{eq:estimacion}), considering $\mathcal{W}_F \approx -Q^{(a)}_\tf{eq}V$ and a small dependence of $\bar{\gamma}$ on $V$, the final velocity grows almost linearly with respect to the bias voltage.
Importantly, in the cases shown in Fig.~\ref{fig:fig4}(a) the final angular velocities fulfill the 
adiabaticity condition $\Omega / \Gamma < k_\tf{B}T/\delta\epsilon$, such that the expansion up to first order in $\Omega$ is adequate in these examples. 

Fig.~\ref{fig:fig4}(b) shows the average value of the instantaneous rotational force (including fluctuations) as function of $\theta$ over $N=4000$ realizations of the time evolution. For the averaging process, we first wait until the rotor arrives to the stationary regime and record the torque within one cycle, i.e. $2\pi n \leq \theta \leq 2 \pi (n+1)$. Obviously, as in each realization the values of $\theta(t)$ are arbitrarily located within this range, to sum the torques obtained from different realizations we group them in a discrete grid of $M = 600$ intervals, i.e. $\theta(t)-2\pi n \rightarrow \theta_k = 2\pi k /M$. If $j$ labels the different realizations, then we have
\begin{equation}
\langle \mathcal{F}_{\theta(t)}^{(i)}+\xi_{\theta(t)} \rangle_N = \frac{1}{N_k} \sum_{j=1}^{N_k} \left( \mathcal{F}_{\theta_k,j}^{(i)}+\xi_{\theta_k,j} \right),
\end{equation}
where $N_k$ counts the number of times $\theta(t)-2\pi n$ fell in the $k$-interval. The figure also shows the standard deviation of the CIF as function of $\theta$, marked as a blue shaded region, which indeed results to be proportional to $\sqrt{\mathcal{D}_\theta}$. It is interesting to note the abrupt profile of $\mathcal{F}_\theta$ and the strong dependence of $\mathcal{D}_\theta$ on $\theta$. While the instantaneous force clearly follows from the double well shape observed in Fig.~\ref{fig:fig3}(a) for trajectory A, the force correlation (and to some extent the current-induced dissipation, due to the fluctuation-dissipation theorem) is zero except in certain narrow regions, associated with transitions between different charge sectors $(n_L,n_R)$.

\begin{figure*}[t]
	\centering
	\includegraphics[width=.9\textwidth]{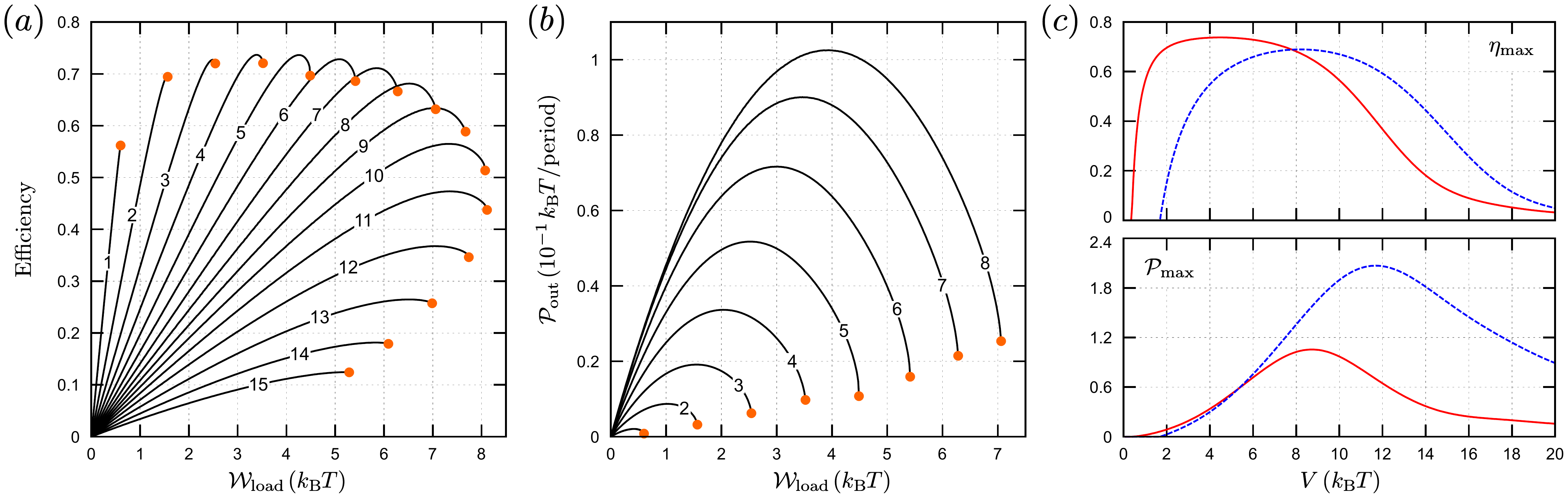}
	\caption{(Color online) (a) Motor efficiency as function of $\mathcal{W}_{\tf{load}}$ for several bias voltages: $V = n~k_\tf{B}T$ 
	with $n=1,2,\ldots,15$ and trajectory A. The orange dots show the limit case $\eta(\mathcal{W}_\tf{load}^{*})$ from which the motor 
	can no longer operate. (b) Output power in trajectory A as function of $\mathcal{W}_\tf{load}$ for the same values of $V$ used in 
	(a). The cases $V > 8~k_\tf{B}T$ start crossing with the other curves and are not shown here to keep the 
	lines distinguishable. The other parameters are the same as those of Fig.~\ref{fig:fig2}. (c) Maximum efficiency (top) and maximum 
	output power (bottom) as function of bias for the trajectories A (solid red) and B (dashed blue) shown in Fig.~\ref{fig:fig2}.}
	\label{fig:fig5}
\end{figure*}

A fundamental concept when investigating devices which perform some effective mechanical work is that of
\textit{efficiency}. Given that the equation of motion of the motor is classical, its meaning will be identical to 
the thermodynamical concept used in conventional motors. In this sense, we define the efficiency 
of this device as the rate $\eta = \mathcal{P}_\tf{out} / \mathcal{P}_\tf{in}$ between output and input 
powers. The input power is given by the amount of energy delivered by the electrons flowing through 
the DQD per period, i.e. $\mathcal{P}_\tf{in} = V (Q^{(i)} + Q^{(a)}) / \tau$. On the other hand, as we 
mentioned before, the amount of useful energy delivered by the motor is $\mathcal{W}_F - \mathcal{E}_\tf{dis}$ 
which, in the stationary regime, coincides with $\mathcal{W}_\tf{load}$. Therefore, the efficiency of the motor can be obtained as
\begin{equation}
	\eta = \frac{\mathcal{W}_F - \mathcal{E}_\tf{dis}}{V(Q^{(i)} + Q^{(a)})}.
	\label{eq:eta}
\end{equation}
In the denominator, the instantaneous contribution $Q^{(i)}$ is related to the induced bias current which, for the considered trajectories, depends on $V$. The adiabatic contribution $Q^{(a)}$, on the other hand, is a constant which only depends on the shape of the trajectory. Notice that in this definition we are not including the force fluctuation, which in general tends to diminish the efficiency, as it increases the average dissipated energy. However, under appropriate conditions, force fluctuations could also enhance $\eta$, as it happens in Brownian motors.~\cite{spiechowicz2014} The role of the force fluctuations in $\eta$ surely deserves further exploration in the regime of transport we are considering here. 

In Fig.~\ref{fig:fig5}(a) we show the motor efficiency as function of $\mathcal{W}_\tf{load}$ for different bias voltages in the range $1\text{--}15$ in units of $k_\tf{B}T$ for trajectory A. In all cases we see that when the loading force is zero, the motor efficiency is simply zero since in the stationary regime all the work done by the motor is dissipated, i.e. $\mathcal{W}_F = \mathcal{E}_\tf{dis}$. For sufficiently small loading forces, $\eta$ grows linearly with a slope which is inversely proportional to $V$, as suggested by Eq.~(\ref{eq:eta}). While increasing $\mathcal{F}_\tf{load}$, however, we need to be careful to avoid crossing the operation condition of the motor 
(orange dots in the figure) since otherwise the motor gets stuck. This can be done by increasing the 
bias voltage: As $\mathcal{W}_F$ is proportional to $V$ and $\mathcal{E}_\tf{dis}$ depends little on 
$V$, the maximum allowed $\mathcal{W}_\tf{load}$ depends linearly on $V$, see Eq.~(\ref{eq:condition1}) and Fig.~\ref{fig:fig3}(b). 
In trajectory A this is true for bias voltages up to $V \simeq 10~k_\tf{B}T$. From this value, the linear dependence of 
$\mathcal{W}_F$ on $V$, as given by Eq.~(\ref{eq:work_bias}) for the linear response regime, no longer holds 
for the chosen trajectory. In fact, for $V \simeq 16~k_\tf{B}T$ the bias deforms the force pseudomagnetic 
field so strongly that $\mathcal{W}_F$ drops even when increasing $V$. As a consequence of this departure from the 
linear regime, for large bias voltages the maximum allowed loading work $\mathcal{W}_\tf{load}^*$ 
decreases, as can be seen in Fig.~\ref{fig:fig3}(b). All this behavior for $\mathcal{W}_\tf{load}$ can be 
easily tracked through the orange dots in Fig.~\ref{fig:fig5}(a). Another point to take into account is that, 
for a fixed bias, the efficiency grows with $\mathcal{W}_\tf{load}$ up to a certain maximum value. This 
maximum is related to the fact that when increasing $\mathcal{F}_\tf{load}$ the rotor slows down 
[see Eq.~(\ref{eq:estimacion})], thus increasing the time employed to complete one cycle and, with it, the 
amount of instantaneous charge $Q^{(i)}$ flowing through the DQD. As the adiabatic charge $Q^{(a)}$ remains 
independent of $\Omega$ (i.e. is a geometric quantity), the denominator in Eq.~(\ref{eq:eta}) grows fast as 
one approaches to the critical point $\mathcal{W}_\tf{load}^*$, meaning that $\mathcal{P}_\tf{out}$ is much 
smaller than $\mathcal{P}_\tf{in}$, the latter dominated by the instantaneous current.

In Fig.~\ref{fig:fig5}(b) we show the output power $\mathcal{P}_\tf{out} = \mathcal{W}_\tf{load}/\tau$ for 
the same cases shown in panel a, up to the bias $V = 8~k_\tf{B}T$, where the curves 
$\mathcal{P}_\tf{out}(\mathcal{W}_\tf{load})$ start decreasing. 
All the curves present a parabolic shape whose maxima locate more or less in the middle of their respective 
allowed ranges for $\mathcal{W}_\tf{load}$. Interestingly, these maxima do not necessarily coincide with 
those of $\eta$. Thus, for a given bias value, one can tune $\mathcal{F}_\tf{load}$ in order to maximize either the efficiency or the output power of the device, but not both.

In Fig.~\ref{fig:fig5}(c) we plot in solid red and dashed blue the maximum efficiencies (upper panel) and output powers (lower panel) as function of the bias voltage, for trajectories A and B, respectively. We can see that the efficiencies are zero up to a finite bias voltage, which marks the transition point from which the energy delivered by the current becomes larger than the amount of energy dissipated by the device, thus ensuring its operation condition. From this critical bias, $\eta_\tf{max}$ suddenly grows up to a plateau, which is sustained up to $V \simeq 6~k_\tf{B}T$ (A) and $V \simeq 10~k_\tf{B}T$ (B). From these values, the maximum efficiency slowly falls to zero. On the other hand, the maximum output power (in both trajectories) does not seem to show these marked regimes as it grows slowly from the critical bias. Rather than a plateau, it shows a peak around $V \simeq 9~k_\tf{B}T$ (A) and $V \simeq 12~k_\tf{B}T$ (B) whose value is $\sim 0.1~k_\tf{B}T$ (A) and $\sim 0.2~k_\tf{B}T$ (B) per period.
Comparing both trajectories, we notice that even in this case where the working points are displayed symmetrically with respect to the symmetry point $\epsilon_0 = k_\tf{B} T \ln(2)-U/2$ (see Fig.~\ref{fig:fig2}), some differences appear in both $\eta_\tf{max}$ and $\mathcal{P}_\tf{max}$. For instance, the maximum efficiency in B starts from a critical bias larger than that of trajectory A, while the maximum output power in B doubles the one obtained in A. As we mentioned before when describing $\mathcal{W}_\tf{eff}$, these differences can be attributed to a stronger conservative part of the CIF along trajectory B, due to a larger average number of particles in the DQD during the cycle. Although the efficiency in all cases does not exceed the value $0.75$, we do not discard greater values in other regimes of the parameters. In any case, this would require some systematic analysis of all the involved parameters which is beyond from this first illustrative example.

\section{Summary and conclusions}\label{sec:conclusions}

We investigated the role of the CIFs in the Coulomb blockade regime within the framework of the real-time diagrammatic approach. On this basis, general expressions were found in the evaluation of the Langevin equation for the dynamics of the slow classical modes. These allowed us to identify the different contributions to the CIFs as: A conservative term related to the Helmholtz's free energy of the local system; a nonconservative contribution that appears in nonequilibrium conditions; a friction term coming from the delayed electronic response to the mechanical motion; and a force fluctuation contribution related with the two-time force correlation function. The expressions were derived assuming quite general conditions: Slow mechanical modes treated classically, perturbative tunnel couplings to the leads, and a local interaction between the electrons and the mechanical degrees of freedom. Therefore, they can be applied to a wide variety of physical problems including, but not exclusively, different forms of nanoelectromechanical devices such as adiabatic quantum motors.
   
At equilibrium conditions, we showed how the Onsager's reciprocity relations and the fluctuation-dissipation 
theorem arise from a real-time diagrammatic treatment. Both proofs emphasize the internal consistency of the obtained expressions for the CIFs and also served to connect them with detail balance ideas. This can be useful to find new ways to break either reciprocity or fluctuation-dissipation relations, and to study their consequences.~\cite{fang2017} Additionally, they provide a physical interpretation for nonconservative CIFs (linked to the pumped current) in terms of auxiliary vector fields and emissivities, thereby opening new perspectives to the study of CIFs in the context of geometric phases.~\cite{yuge2012,pluecker2017}  We should also mention that the proven Onsager's relations rely on a general scheme that could be used in other quantities (e.g. heat and spin currents) as far as their linear response coefficients admit the form given by Eq.~(\ref{eq:Lcoeff}).

To illustrate the obtained general expressions for the CIFs, we considered a double quantum dot based motor. Here, we analyzed its operation conditions as function of several parameters including the bias voltage, the moment of inertia, the loading force, as well as the mechanical working point (see Fig.~\ref{fig:fig2}).
When doing so, we derived a simple and efficient recursive formula (see App.~\ref{app:4}) that allows one to predict under which conditions the motor will operate as such. The method can be used in place of the explicit time integration of the equation of motion while still providing the position-dependent steady-state velocity of the motor with high accuracy. Although we did not perform an exhaustive exploration in the space of parameters, we were able to obtain maximum efficiencies up to 0.75. Comparing these values with those obtained in the open quantum dot example of Ref.~\onlinecite{fernandez2015}, the Coulomb blockade regime seems promising for the design of highly-efficient adiabatic quantum motors.

To explore the role of the Coulomb interaction in CIFs, we analyzed two different charge regions in the DQD stability diagram, characterized by $0 \leftrightarrow 1$ and $1 \leftrightarrow 2$ charge transitions. This was accomplished by considering two possible trajectories, each one centered around a triple point (see Fig.~\ref{fig:fig2}). We found strong differences in the perfomance (efficiency and output power) as one changes the motor's operational region [see Fig.~\ref{fig:fig5}(c)], due to the role of the conservative part of the CIF in each case. This result is surprising to some extent, as the only difference occurring in quantum pumping is essentially a change of sign in the pumped current.~\cite{riwar2010,juergens2013}

We believe this work paves the way to further investigations on CIFs in quantum devices dominated by strong Coulomb interactions and weakly coupled to the leads. In particular, it would be interesting to extend the obtained formulas to nonlocal forces as well as to higher-orders terms in both the tunnel coupling and the modulation frequency.

\vspace{0.5cm}

\noindent
\textit{Acknowlegdments.--} This work was supported by Consejo Nacional de Investigaciones Cient\'ificas y 
T\'ecnicas (CONICET), Secretar\'ia de Ciencia y Tecnolog\'ia -- Universidad Nacional de C\'ordoba (SECYT--UNC), 
and Ministerio de Ciencia y Tecnolog\'ia  de la Provincia de C\'ordoba (MINCyT--Cba). All authors are members 
of CONICET.

\appendix
\section{Coordinate dependence in eigenstates}
\label{app:1}

Here we discuss why any $\bm{X}$-dependence in the local system's eigenstates can be disregarded in the CIFs, as far as the off-diagonal elements of the reduced density operator are decoupled from the diagonal ones to lowest order in $\Gamma$.
For the present purpose, let us assume the following form for the local system Hamiltonian
\begin{equation}
\hat{H}_\tf{el}(\bm{X}) = \sum_\alpha E_\alpha(\bm{X}) \ket{\alpha(\bm{X})} \bra{\alpha(\bm{X})},
\end{equation}
where we take $\bm{X}$ as a set of classical variables. The force operator, defined as the $\bm{X}$-gradient of the local Hamiltonian, i.e. $\hat{\bm{F}} = -\hat{\nabla H}_\tf{el}$, takes the form:
\begin{equation}
\hat{\bm{F}} = -\sum_\alpha \left\{ \nabla E_\alpha \ket{\alpha} \bra{\alpha} + E_\alpha \left[ (\nabla\ket{\alpha}) \bra{\alpha} + \ket{\alpha} (\nabla\bra{\alpha}) \right] \right\},
\end{equation}
where we skip the $\bm{X}$-arguments in all quantities to keep the notation simple. If we now evaluate the 
matrix elements of the force operator in the $\bm{X}$-eigenbasis, we obtain
\begin{equation}
\bm{F}_{\alpha \beta} = \bra{\alpha}\hat{\bm{F}}\ket{\beta} = - \nabla E_\alpha \delta_{\alpha \beta} - (E_\beta-E_\alpha) \bra{\alpha}\nabla\ket{\beta},
\end{equation}
where we used $\nabla(\braket{\alpha|\beta}) = 0$. 
The above equation therefore suggests that if there is some explicit $\bm{X}$-dependence in the eigenstates, then it could contribute in the force as an off-diagonal element. By tracing $\hat{\bm{F}}$ with the instantaneous (or adiabatic) reduced density operator $\hat{p}^{(i/a)}$ we obtain
\begin{equation}
\langle \hat{\bm{F}} \rangle^{(i/a)}  =  - \sum_\alpha \nabla E_\alpha p_{\alpha \alpha}^{(i/a)} - \sum_{\alpha \beta} (E_\beta-E_\alpha) 
\bra{\alpha}\nabla\ket{\beta} p_{\beta \alpha}^{(i/a)}.
\end{equation}
Clearly, the  contribution from the $\bm{X}$-dependence in the eigenstates only appears through the off-diagonal elements $p_{\alpha\beta}$ of the reduced density operator. However, when $|E_\alpha - E_\beta| \gg \Gamma$ or the involved states in $p_{\alpha\beta}$ differ in charge or spin, the dynamics of the off-diagonal elements decouple from those of the diagonal ones to lowest order in $\Gamma$, meaning that coherent effects due to $p_{\alpha\beta}$ can be disregarded on this level of approximation.~\cite{leijnse2008,reckermann_phd}

Notice that in the example discussed in Sec.~\ref{sec:dqd} the coherences 
$\bra{\tf{b} \sigma} \hat{p} \ket{\tf{a} \sigma}$ and $\bra{\tf{a} \sigma} \hat{p} \ket{\tf{b} \sigma}$ could in principle be coupled with the occupations since their involved states belong to the same charge and spin 
sectors. In fact, these need to be taken into account in the 
weak interdot coupling regime where $t_c \lesssim \Gamma$, and are responsible for level 
renormalizations in both the instantaneous and adiabatic charge currents.~\cite{wunsch2005,riwar2010} In our 
case, however, we consider a strong interdot coupling regime where such effects can be disregarded to lowest order in $\Gamma$.

\section{Auxiliary formulas for reciprocity relations}
\label{app:2}

In this appendix we derive the general expressions proposed in Eqs.~(\ref{eq:ons_gral}), (\ref{eq:occ_der1}) 
and (\ref{eq:occ_der2}) which allow us to prove all the reciprocity relations discussed in Sec.~\ref{subsec:onsager} 
between the current induced force and the charge tunnel current in equilibrium.

\textit{Symmetry relation -} Let us begin with the demonstration of Eq.~(\ref{eq:ons_gral}) for two arbitrary observables 
$A$ and $B$. For an arbitrary observable $B$, with associated kernel $\bm{W}^B$ and response coefficients
\begin{equation}
\varphi_\alpha^B = \sum_{\beta \gamma} W_{\beta \gamma}^B \tilde{W}_{\gamma \alpha}^{-1},
\end{equation}
we want to prove that the following expression
\begin{align}
\mathcal{J}_{AB} &\equiv \sum_{\alpha \beta} W_{\alpha \beta}^A \left( \varphi_{\beta}^B - \bar{\varphi}^B \right) p_\beta^{(i)} \nonumber \\
&= \sum_{\alpha \beta} \sum_{\gamma_1 \gamma_2} W_{\alpha \beta}^A W_{\gamma_1 \gamma_2}^B \sum_\kappa \left( \tilde{W}_{\gamma_2 \beta}^{-1} 
- \tilde{W}_{\gamma_2 \kappa}^{-1} \right) p_\beta^{(i)} p_\kappa^{(i)},
\end{align}
is invariant under interchange of $A$ and $B$ observables when evaluated in equilibrium, i.e. $\mathcal{J}_{AB,\tf{eq}} = \mathcal{J}_{BA,\tf{eq}}$. This implies that the following relation must hold in equilibrium:
\begin{equation}
\sum_\kappa \left( \tilde{W}_{\alpha \beta}^{-1} - \tilde{W}_{\alpha \kappa}^{-1} \right) p_\beta ^{(i)} p_\kappa^{(i)} =
\sum_\kappa \left( \tilde{W}_{\beta \alpha}^{-1} - \tilde{W}_{\beta  \kappa}^{-1} \right) p_\alpha^{(i)} p_\kappa^{(i)}.
\label{eq:invdetbal}
\end{equation}
In order to prove the above relation, we use the detailed balance property of the 
evolution kernel in equilibrium, which reads $W_{\alpha \beta} p_\beta^{(i)} = W_{\beta \alpha} p_\alpha^{(i)}$. 
Provided that $\bm{W}\bm{p}^{(i)} = 0$, this relation can be extended to the (invertible) kernel as
\begin{equation}
\sum_\kappa \left( \tilde{W}_{\alpha \beta} - \tilde{W}_{\alpha \kappa} \right) p_\beta ^{(i)} p_\kappa^{(i)} =
\sum_\kappa \left( \tilde{W}_{\beta \alpha} - \tilde{W}_{\beta  \kappa} \right) p_\alpha^{(i)} p_\kappa^{(i)}.
\label{eq:detbal}
\end{equation}
The similarity between Eqs.~(\ref{eq:invdetbal}) and (\ref{eq:detbal}) suggests that this relation holds for any power 
of the kernels. Therefore, we now test the above relation for $\tilde{\bm{W}}^{n}$, 
with $n=1,2,...$. Let us define
\begin{align}
\mathcal{J}^{(n)}_1 &= \sum_\kappa \left( \tilde{W}_{\alpha \beta}^n - \tilde{W}_{\alpha \kappa}^n \right) p_\beta ^{(i)} p_\kappa^{(i)}, \\
\mathcal{J}^{(n)}_2 &= \sum_\kappa \left( \tilde{W}_{\beta \alpha}^n - \tilde{W}_{\beta  \kappa}^n \right) p_\alpha^{(i)} p_\kappa^{(i)}.
\end{align}  
By induction, if we now suppose that $\mathcal{J}^{(n)}_1 = \mathcal{J}^{(n)}_2$, then for $n+1$ we have
\begin{align}
\mathcal{J}^{(n+1)}_1 
&= \sum_{\kappa\gamma} \tilde{W}_{\alpha\gamma} \left(\tilde{W}^n_{\beta\gamma}-\tilde{W}^n_{\beta\kappa}\right) p_\gamma^{(i)} p_\kappa^{(i)}, \\
\mathcal{J}^{(n+1)}_2 
&= \sum_{\kappa\gamma} \tilde{W}^n_{\beta \gamma} \left(\tilde{W}_{\alpha\gamma}-\tilde{W}_{\alpha\kappa}\right) p_\gamma^{(i)} p_\kappa^{(i)},
\end{align}
and since the indices $\kappa$ and $\gamma$ run over all possible eigenstates of the local system, we obtain that $\mathcal{J}^{(n+1)}_1 = \mathcal{J}^{(n+1)}_2$. Noticing that the pseudoinverse kernel can be written as $\tilde{\bm{W}}^{-1} = \sum_n c_n \tilde{\bm{W}}^n$, we prove Eq.~(\ref{eq:invdetbal}).

\textit{Occupation derivatives -} We now begin with the $\mu_r$-derivative of the instantaneous occupations in the local system. Our starting point 
is the instantaneous kinetic equation Eq.~(\ref{eq:kineqin}), which after derivation with respect 
to $\mu_r$ reads:
\begin{equation}
\bm{W} \dd{\bm{p}^{(i)}}{\mu_r} = - \dd{\bm{W}}{\mu_r} \bm{p}^{(i)}.
\end{equation}
Taking matrix elements with respect to the diagonal basis, we obtain
\begin{equation}
\sum_\beta W_{\alpha \beta} \dd{p_\beta^{(i)}}{\mu_r} = -\sum_\beta \dd{W_{\alpha \beta}}{\mu_r} p_\beta^{(i)}.
\end{equation}
The next step is to separate the evolution kernel in diagonal and off-diagonal parts, i.e. $\bm{W} = 
\bm{W}^\mathrm{d} + \bm{W}^\mathrm{n}$, such that the above reads
\begin{align*}
\sum_\beta W_{\alpha \beta} \dd{p_\beta^{(i)}}{\mu_r} &= -\sum_\beta \left( \dd{W_{\alpha \beta}^\tf{d}}{\mu_r} + \dd{W_{\alpha \beta}^\tf{n}}{\mu_r} \right) p_\beta^{(i)} \\
	&= \sum_\beta \left( \dd{W_{\beta \alpha}^\tf{n}}{\mu_r} p_\alpha^{(i)} - \dd{W_{\alpha \beta}^\tf{n}}{\mu_r} p_\beta^{(i)} \right) \\
	&= \frac{1}{k_\tf{B}T} \sum_\beta f_{\alpha \beta}^r f_{\beta \alpha}^r \left( \eta_{\beta \alpha} \Gamma_{\beta \alpha}^r p_\alpha^{(i)} - \eta_{\alpha \beta} \Gamma_{\alpha \beta}^r p_\beta^{(i)} \right).
\end{align*}
In the above steps, we used that for the instantaneous kernel $W_{\alpha \beta}^\mathrm{d} = 
W_{\alpha \alpha} \delta_{\alpha \beta} = -\sum_\gamma W_{\gamma \alpha}^\mathrm{n} \delta_{\alpha \beta}$ and 
the explicit form $W_{\alpha \beta}^\mathrm{n} = \sum_r \Gamma_{\alpha \beta}^r f_{\alpha \beta}^r$, where
\begin{equation}
f_{\alpha \beta}^r = \frac{1}{1+\exp \left[(E_\alpha-E_\beta-\eta_{\alpha \beta}\mu_r)/k_\tf{B}T \right]},
\end{equation}
and $\eta_{\alpha \beta} = n_\alpha - n_\beta = -\eta_{\beta \alpha}$ indicates 
whether the local system gains or loses one electron after the tunnel event. The derivative of the kernel matrix element therefore reads
\begin{equation}
\dd{W_{\alpha \beta}^\tf{n}}{\mu_r} = \frac{1}{k_\tf{B}T} \Gamma_{\alpha \beta}^r \eta_{\alpha \beta} f_{\alpha \beta}^r \, f_{\beta \alpha}^r.
\end{equation}
We now consider the equilibrium condition $\mu_L = \mu_R = \mu$ for the reservoirs. We here simplify this 
condition by setting $\mu$ as the reference origin for the addition energies, i.e. $\mu = 0$. Therefore, in equilibrium one obtains
\begin{equation}
f_{\alpha \beta}^r \rightarrow f_{\alpha \beta} = \frac{1}{1+\exp\left[(E_\alpha - E_\beta)/k_\tf{B}T \right]}.
\end{equation}
The following assumption relies on the symmetry property for the tunnel processes, i.e. 
$\Gamma_{\alpha \beta}^r = \Gamma_{\beta \alpha}^r$, and hence we have
\begin{equation}
\sum_\beta W_{\alpha \beta} \dd{p_\beta^{(i)}}{\mu_r} = -\frac{1}{k_\tf{B}T} \sum_\beta \Gamma_{\alpha \beta}^r \eta_{\alpha \beta} f_{\alpha \beta} f_{\beta \alpha}
(p_\alpha^{(i)} + p_\beta^{(i)}).
\end{equation}
Additionally, since we are now in equilibrium, the occupations are described through 
Boltzmann factors, i.e.
\begin{equation}
p_\alpha^{(i)} = \frac{\exp(- E_\alpha/k_\tf{B}T)}{\sum_\beta \exp(- E_\beta/k_\tf{B}T)},
\label{eq:boltzmann}
\end{equation} 
and hence
\begin{eqnarray}
\sum_\beta W_{\alpha \beta} \dd{p_\beta^{(i)}}{\mu_r} &=& -\frac{1}{k_\tf{B}T} \sum_\beta \Gamma_{\alpha \beta}^r \eta_{\alpha \beta} f_{\alpha \beta} p_\beta^{(i)} \\
&=& -\frac{1}{k_\tf{B}T} \sum_\beta \Gamma_{\alpha \beta}^r \eta_{\alpha \beta} f_{\beta \alpha} p_\alpha^{(i)},
\end{eqnarray}
where we used that $f_{\alpha \beta}p_\beta^{(i)} = f_{\beta \alpha}p_\alpha^{(i)}$. We now use the following 
property for the charge current kernel to lowest order, i.e. $W^{I_r}_{\alpha \beta} = -\eta_{\alpha\beta} W^r_{\alpha \beta}$, such that the above equations can be written as
\begin{eqnarray}
\sum_\beta W_{\alpha \beta} \dd{p_\beta^{(i)}}{\mu_r} &=& +\frac{1}{k_\tf{B}T} \sum_\beta W^{I_r}_{\alpha \beta} p_\beta^{(i)} \label{eq:a1}\\
&=& -\frac{1}{k_\tf{B}T} \sum_\beta W^{I_r}_{\beta \alpha} p_\alpha^{(i)}.
\label{eq:a2}
\end{eqnarray}
The above allows us to write the derivative of the occupations vector as
\begin{equation}
\dd{\bm{p}^{(i)}}{\mu_r} = \frac{1}{k_\tf{B}T} \tilde{\bm{W}}^{-1} \bm{W}^{I_r} \bm{p}^{(i)},
\end{equation}
so that
\begin{equation}
\dd{p_\alpha^{(i)}}{\mu_r} = \frac{1}{k_\tf{B}T} \sum_{\gamma_1 \gamma_2} \tilde{W}^{-1}_{\alpha \gamma_1} W^{I_r}_{\gamma_1 \gamma_2} p_{\gamma_2}^{(i)},
\end{equation}
and using Eqs.~(\ref{eq:a1}) and (\ref{eq:a2}) we obtain
\begin{eqnarray*}
\dd{p_\alpha^{(i)}}{\mu_r} &=& -\frac{1}{k_\tf{B}T} \sum_{\gamma_1 \gamma_2} W^{I_r}_{\gamma_2 \gamma_1} \tilde{W}^{-1}_{\alpha \gamma_1} p_{\gamma_1}^{(i)} \\
&=& -\frac{1}{k_\tf{B}T} \sum_{\gamma_1 \gamma_2} W^{I_r}_{\gamma_2 \gamma_1} \sum_\beta (\tilde{W}^{-1}_{\alpha \gamma_1} - \tilde{W}^{-1}_{\alpha \beta}) p_\beta^{(i)}  p_{\gamma_1}^{(i)},
\end{eqnarray*}
where we used $\sum_\beta p_\beta^{(i)} = 1$ and that the instantaneous current in equilibrium is 
zero, i.e. $I_r^{(i)} = \sum_{\gamma_1 \gamma_2 } W^{I_r}_{\gamma_2 \gamma_1} p_{\gamma_1}^{(i)} = 0$, and 
thus the second term in the above equation is simply zero. Using Eq.~(\ref{eq:invdetbal}) allows us to interchange the subindices and obtain the 
proposed expression for the occupation derivatives in Eq.~(\ref{eq:occ_der1}), which explicitely reads as
\begin{equation}
\dd{p_\alpha^{(i)}}{\mu_r} = -\frac{1}{k_\tf{B}T} \sum_{\gamma_1 \gamma_2} W^{I_r}_{\gamma_2 \gamma_1} 
\sum_\beta (\tilde{W}^{-1}_{\gamma_1 \alpha} - \tilde{W}^{-1}_{\gamma_1 \beta}) p_\beta^{(i)}  p_\alpha^{(i)}.
\end{equation}

We now continue with Eq.~(\ref{eq:occ_der2}) for the $\dot{X}_\nu$-derivative of the adiabatic occupations in
equilibrium. From Eq.~(\ref{eq:occ_ad}) we have
\begin{equation}
\dd{p_\alpha^{(a)}}{\dot{X}_\nu} = \sum_\beta \tilde{W}_{\alpha \beta}^{-1}\dd{p_\beta^{(i)}}{X_\nu},
\end{equation}
and using Eq.~(\ref{eq:boltzmann}), we obtain
\begin{equation}
\dd{p_\alpha^{(a)}}{\dot{X}_\nu} = \frac{1}{k_\tf{B}T} \sum_\beta \tilde{W}_{\alpha \beta}^{-1}\left(F_{\nu,\beta} - F_\nu^{(i)} \right) p_\beta^{(i)}, \label{eq:der1}
\end{equation}
where $F_{\nu,\beta} = - \partial E_\beta / \partial X_\nu$ is the $\beta$-element of the $\nu$-component of
the force operator and $F_\nu^{(i)} = \sum_\beta F_{\nu,\beta} p_\beta^{(i)}$ is the instantaneous force.
The proposed expression in Eq.~(\ref{eq:occ_der2}) is:
\begin{equation}
\dd{p_\alpha^{(a)}}{\dot{X}_\nu} = \frac{1}{k_\tf{B}T} \left( \varphi_\alpha^{F_\nu} - \bar{\varphi}^{F_\nu} \right) p_\alpha^{(i)}.
\label{eq:der2}
\end{equation}
So now we should arrive to Eq.~(\ref{eq:der1}) from Eq.~(\ref{eq:der2}). To do so, we use the definition of 
the force response coefficients in terms of the pseudoinverse kernel, which yields
\begin{equation}
\dd{p_\alpha^{(a)}}{\dot{X}_\nu} = \frac{1}{k_\tf{B}T} \sum_\beta F_{\nu,\beta} \sum_\gamma 
\left( \tilde{W}_{\beta \alpha}^{-1} - \tilde{W}_{\beta \gamma}^{-1} \right) p_\gamma^{(i)} p_\alpha^{(i)}.
\end{equation}
By using Eq.~(\ref{eq:invdetbal}) we can rewrite the second sum and obtain
\begin{equation}
\dd{p_\alpha^{(a)}}{\dot{X}_\nu} = \frac{1}{k_\tf{B}T} \sum_\beta F_{\nu,\beta} \sum_\gamma 
\left( \tilde{W}_{\alpha \beta}^{-1} - \tilde{W}_{\alpha \gamma}^{-1} \right) p_\gamma^{(i)} p_\beta^{(i)},
\end{equation}
such that combining the two sums we arrive to Eq.~(\ref{eq:der1}).

\section{Local system's correlation function}
\label{app:3}

In this section we derive, along the lines of the real-time diagrammatic approach discussed in Refs.~\onlinecite{thielmann2003,riwar2013,riwar_phd}, the time-dependent correlation function for the fluctuation of two \textit{local} observables $A$ and $B$, namely
\begin{align}
D_{AB}(t) &= \int_{-\infty}^{\infty} dt' D_{AB}(t,t') \label{eq:corr} \\
			 &= \frac{1}{2} \int_{-\infty}^{\infty} dt' \left[ \langle \hat{\xi}_A(t) \hat{\xi}_B(t') \rangle + \langle \hat{\xi}_B(t') \hat{\xi}_A(t) \rangle \right],
\end{align}
where $\hat{\xi}_A(t) = \hat{A}(t) - \langle \hat{A}(t) \rangle$ is the $A$-fluctuation operator in the Heisenberg picture, such that
$\hat{A}(t) = \hat{U}(t_0,t) \hat{A}_t \hat{U}(t,t_0)$, with $\hat{U}(t,t_0) = \mathcal{T} \exp \left[-i \int_{t_0}^t \hat{H}(\tau) d\tau \right]$ the full system's propagator and $\hat{A}_t$ the 
local operator written in the Schr\"{o}dinger picture, which parametrically depends on time through the mechanical coordinate $\bm{X}(t)$. The 
two-time correlation function $D_{AB}(t,t')$ then takes the form:
\begin{equation}
D_{AB}(t,t') = \frac{1}{2} \langle \{ \hat{A}(t),\hat{B}(t') \} \rangle - \langle \hat{A}(t) \rangle \langle \hat{B}(t') \rangle,
\label{eq:twotime}
\end{equation}
with $\{ \bullet \, , \bullet \}$ the anticommutator. Let us begin with the term carrying the anticommutator, i.e. $\mathcal{C}_{AB} = \langle \{ \hat{A}(t),\hat{B}(t') \} \rangle$, that can be written as:
\begin{equation}
\mathcal{C}_{AB} = \frac{1}{2} \tr{} \, \{ \hat{A}(t), \{ \hat{B}(t'), \hat{\rho}_0 \} \}, 
\end{equation}
where the trace involves all electronic degrees of freedom, including both the local system and the reservoirs. The full density matrix $\hat{\rho}$ is evaluated at the initial time $t_0$, from which the local system and the reservoirs are assumed to be coupled adiabatically. Under this assumption, the density matrix at $t_0$ can be factorized as $\hat{\rho}_0 = \hat{\rho}_\tf{res} \hat{p}_0$, where $\hat{\rho}_\tf{res}$ is the density operator of the reservoirs (assumed to be always in equilibrium) and $\hat{p}_0 = \hat{p}(t_0)$ is the density operator of the local system. By defining the superoperator $L^A(t) \, \bullet = \{ \hat{A}(t), \bullet \}$, we obtain
\begin{equation}
\mathcal{C}_{AB} = \frac{1}{2} \tr{} \, L^A(t) L^B(t') \hat{\rho}_0.
\end{equation}
We now write the above superoperators in the Schr\"{o}dinger picture, i.e. $L_A(t) = \pi(t_0,t) L^A_t \pi(t,t_0)$, where $\pi(t,t_0) = \mathcal{T} \exp \left[-i \int_{t_0}^t L(\tau) d\tau \right]$ represents the (superoperator) propagator of the full system. 
Here $L(t) \, \bullet = [\hat{H}(t),\bullet]$, with $[\bullet \, , \bullet ]$ denoting commutation, is the full system's Liouvillian superoperator. Replacing these expressions we arrive to
\begin{equation}
\mathcal{C}_{AB} = \frac{1}{2} \tr{} \, L^A_t \pi(t,t')  L^B_{t'} \pi(t',t_0) \hat{\rho}_0,
\end{equation}
where we use that the leftmost propagator, $\pi(t_0,t)$, can only act as the identity due to the invariance of the trace under cyclic permutations. Importantly, as the time integral in Eq.~(\ref{eq:corr}) involves the cases $t'<t$ and $t'>t$, we can rearrange the above superoperators in chronological order as follows
\begin{equation}
\mathcal{C}_{AB} = \frac{1}{2} \tr{} \left\{
\begin{array}{lr}
L^A_t \pi(t,t')  L^B_{t'} \pi(t',t_0) \hat{\rho}_0, & t'<t \\
L^B_{t'} \pi(t',t)  L^A_t \pi(t,t_0) \hat{\rho}_0, & t'>t
\end{array}\right. \label{eq:start}
\end{equation}
Indeed, this equation is a general expression for the two-time correlation function in the sense that nothing was said yet about the local nature of the involved observables $A$ and $B$. In fact, the same expression was used as a starting point for the current noise,~\cite{riwar_phd} provided the above local superoperators are replaced by non-local ones, related to the charge current flowing from/into the leads. The main difference in our case is that the force operator consists of local system's field operators only, while the current operator is bilinear in the local system and reservoirs, similar to the tunnel Hamiltonian. This radical difference implies a different diagrammatic treatment as compared to the current noise, in the sense that, here, the superoperators $L^A$ and $L^B$ cannot be considered on the same level than the tunnel Liouvillian. Technically speaking, the local superoperators are not external vertices to be contracted. 

\begin{figure}[h]
	\centering
	\includegraphics[width=\columnwidth]{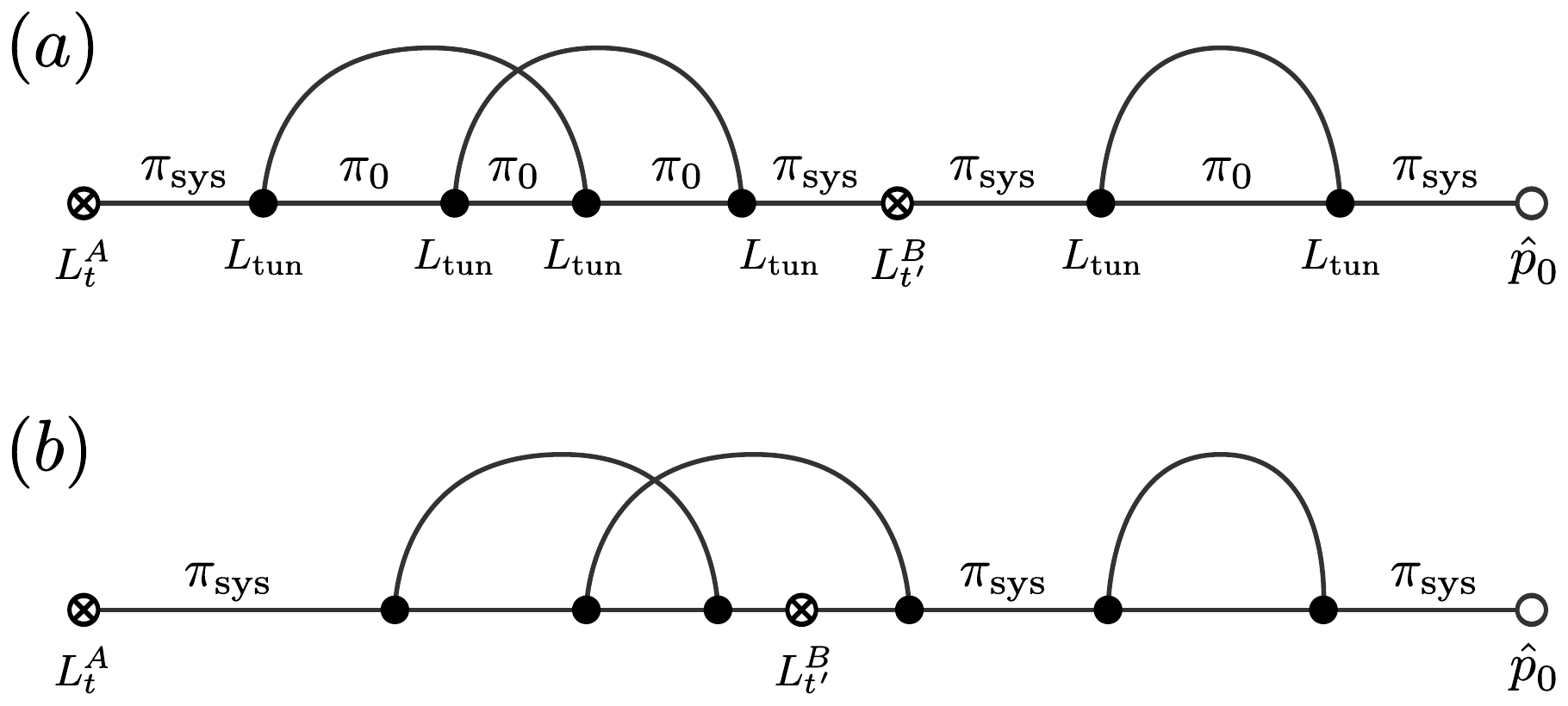}
	\caption{Examples of diagrams contributing to $\mathcal{C}_{AB}$. (a) Contribution to Eq.~(\ref{eq:case1}) where the local superoperators are not 		contained within an irreducible block. (b) Contribution to Eq.~(\ref{eq:case2}) where one of the local superoperators is contained within an irreducible block.}
	\label{fig:fig6}
\end{figure}

In order to treat Eq.~(\ref{eq:start}) diagrammatically, we consider the Dyson equation for the full propagator
\begin{equation}
\pi(t,t') = \pi_0(t,t')+(-i)\int_{t'}^t dt_1 \pi_0(t,t_1)L_\tf{tun}(t_1) \pi(t_1,t'),
\end{equation}
where $\pi_0(t,t')=\mathcal{T} \exp \left[-i\int_{t'}^t L_0(\tau) d\tau \right]$ is the propagator of the decoupled system, defined by $L_0(t) \, \bullet = \left[ \hat{H}_\tf{sys}(t) + \hat{H}_\tf{res} , \bullet \right]$. Since the local superoperators in Eq.~(\ref{eq:start}) do not contain reservoir's field operators, and given that there is a trace over the reservoir's degrees of freedom, the total number of tunnel Liouvillians needs to be always even. With this in mind, and the expansion of the above Dyson equation, we can construct different diagrams contributing to $\mathcal{C}_{AB}$. In Fig.~\ref{fig:fig6} we show two examples of diagrams for the case $t'<t$. Here, the black dots (vertices) represent tunnel Liouvillians evaluated at different times, while the crossed circles denote the local superoperators. The tunnel Liouvillians are connected through contraction lines involving the trace over the reservoir's degrees of freedom and the horizontal lines correspond to a free propagation in the local system, i.e. $\pi_\tf{sys}(t,t')=\mathcal{T} \exp \left[-i\int_{t'}^t L_\tf{sys}(\tau) d\tau \right]$.

A diagram contributing to $\mathcal{C}_{AB}$ thus consists of a series of irreducible blocks connected by a free propagation. By irreducible block we mean those regions where any vertical cut intersects with at least one contraction line. The shown diagrams, in fact, belong to two different types of contributions one can find when constructing $\mathcal{C}_{AB}$:  

1) When all irreducible blocks do not contain a local superoperator [see Fig.~\ref{fig:fig6}(a)]. In this case we have that the sum of all possible diagrams yields the following contribution to $\mathcal{C}_{AB}$
\begin{equation}
\frac{1}{2} e^\tf{T} \left\{
\begin{array}{lr}
L^A_{t } \Pi(t ,t') L^B_{t'} \Pi(t',t_0) \hat{p}_0, & t'<t \\
L^B_{t'} \Pi(t',t ) L^A_{t } \Pi(t ,t_0) \hat{p}_0, & t'>t
\end{array}\right. \label{eq:case1}
\end{equation}
where $\Pi(t,t') = \tf{tr}_\tf{res} \, \pi(t,t')\hat{\rho}_0$ is the local system's reduced propagator and $e^\tf{T}$ represents the trace over the local system degrees of freedom.

2) When one of the local superoperators is contained within an irreducible block [see Fig.~\ref{fig:fig6}(b)]. Notice that since the local superoperators do not act as contraction vertices, only the earliest superoperator can follow this rule. By identifying this irreducible block as an operator-related kernel we obtain that all diagrams of this type add up to yield the following contribution
\begin{equation}
\frac{1}{2} e^\tf{T} \left\{
\begin{array}{lr}
\int_{t'}^t dt_1 \int_{t_0}^{t'} dt_2 L^A_{t } \Pi(t ,t_1) K^B_{t'}(t_1,t_2) \Pi(t_2,t_0) \hat{p}_0, & t'<t \\
\int_t^{t'} dt_1 \int_{t_0}^{t } dt_2 L^B_{t'} \Pi(t',t_1) K^A_{t }(t_1,t_2) \Pi(t_2,t_0) \hat{p}_0, & t'>t
\end{array}\right. \label{eq:case2}
\end{equation}
Here $K^A_t(t_1,t_2)$ and $K^B_{t'}(t_1,t_2)$ are the local observable related kernels, which in $k$-th order in the tunnel coupling strength $\Gamma$ can be written in general as
\begin{widetext}
\begin{equation}
K^{A \, (2k)}_t(t_f,t_i) = \sum_\tf{irr. contr.} \sum_\tf{perm} \int_{t_i}^{t_f} \widehat{dt} \,
\tr{res} \, L_\tf{tun}(t_f) \pi_0(t_f,t_{2k-1}) \ldots L_\tf{tun}(t_n)\pi_0(t_n,t)L^A_t\pi_0(t,t_{n-1}) \ldots 
L_\tf{tun}(t_2)\pi_0(t_2,t_i)L_\tf{tun}(t_i)\hat{\rho}_\tf{res},
\label{eq:kernelA}
\end{equation}
\end{widetext}
where the sums mean that only irreducible contractions and all possible positions for the local superoperator $L^A_t$ (provided is surrounded by two tunnel Liouvillians) need to be taken into account. The time-integral symbol is a shortcut for the time ordered integrals:
\begin{equation}
\int_{t_i}^{t_f} \widehat{dt} \rightarrow \int_{t_i}^{t_f} dt_{2k-1} \int_{t_i}^{t_{2k-1}} dt_{2k-2} \ldots \int_{t_i}^{t_4} dt_3 \int_{t_i}^{t_3} dt_2.
\end{equation} 
For $K^{B}_{t'}$ we can simply replace in Eq.~(\ref{eq:kernelA}) the local superoperator $L^A_t$ by $L^B_{t'}$. Notice that the lowest order kernel is necessarily linear in $\Gamma$, such that in a lowest order calculation these type of contributions can be disregarded against those appearing in case 1.

The remaining term in Eq.~(\ref{eq:twotime}) corresponds to the mean values of the local observables $A$ and $B$ at times $t$ and $t'$, respectively. They can be simply written as:
\begin{equation}
\langle \hat{A}(t) \rangle \langle \hat{B}(t') \rangle = \frac{1}{4} e^\tf{T} L^A_t \Pi(t,t_0) \hat{p}_0 \otimes e^\tf{T} L^B_{t'} \Pi(t',t_0) \hat{p}_0,
\end{equation}
such that the lowest order contribution to the two-time correlation function can be written as
\begin{equation}
D_{AB}(t,t') = \frac{1}{4} e^\tf{T} \left\{
\begin{array}{lr}
L^A_{t } \bar{\Pi}(t,t') L^B_{t'}\hat{p}(t'), & t'<t \\
L^B_{t'} \bar{\Pi}(t',t) L^A_{t }\hat{p}(t ), & t'>t
\end{array}\right.
\end{equation}
with $\bar{\Pi}(t_1,t_2) = \Pi(t_1,t_2) - \hat{p}(t_1) \otimes e^\tf{T}$. With this result for $D_{AB}(t,t')$ we can now integrate over $t'$ and write the correlation function as
\begin{align}
D_{AB}(t) ={} &  \frac{1}{4} \int_{-\infty}^t dt' e^\tf{T} L^A_{t } \bar{\Pi}(t,t') L^B_{t'} \hat{p}(t') \nonumber \\
			 	  &+ \frac{1}{4} \int_t^{ \infty} dt' e^\tf{T} L^B_{t'} \bar{\Pi}(t',t) L^A_{t } \hat{p}(t).
\end{align}
Since we are interested in the instantaneous (i.e. zeroth order in $\Omega$) and lowest order in $\Gamma$ contributions, the two local 
superoperators can be evaluated at time $t$ while $\bar{\Pi}(t,t') \rightarrow \bar{\Pi}^{(i)}_t(t-t')$. As we already mentioned, the subindex $t$ indicates a parametric dependence on $t$ due to the mechanical coordinate $\bm{X}(t)$. The integral over $t'$ can thus be taken as the zero frequency Laplace transform and we obtain
\begin{equation}
D_{AB}(t) = \frac{e^\tf{T}}{4} L^A_t \bar{\Pi}^{(i,-1)}_t L^B_t \hat{p}^{(i,0)}_t + (A \leftrightarrow B),
\end{equation}
where $\bar{\Pi}^{(i,-1)}_t = [\tilde{W}^{(i,1)}_t]^{-1}(\hat{p}^{(i,0)}_t \otimes e^\tf{T}-1)$ and $\tilde{W}^{(i,1)}_t$ is the pseudo invertible kernel defined after Eq.~(\ref{eq:occ_ad}). We refer to Refs.~\onlinecite{riwar2013,riwar_phd} for more details on the calculation of the different orders of $\bar{\Pi}$. The symbol $A \leftrightarrow B$ means that the second term (due to $t'>t$) writes as the first one but with $A$ and $B$ exchanged.

Now that we have the general expression for the correlation function of two local observables, we can replace them by the different components of the current induced force. In doing so, we can formally define a local kernel in time-domain as
\begin{equation}
W^A(t,t') = \frac{1}{2} L^A \Pi(t,t') \delta(t-t'),
\end{equation}
such that its zero-frequency Laplace transform simply reads as $W^A \, \bullet = \{\hat{A},\bullet \}/2$. The matrix elements of this superoperator are given by~\cite{leijnse2008}
\begin{equation}
[W^A]^{b_{+}b_{-}}_{a_{+}a_{-}} = \frac{1}{2} \left( \bra{a_{+}}\hat{A}\ket{b_{+}}\delta_{a_{-}b_{-}}+\bra{b_{-}}\hat{A}\ket{a_{-}}\delta_{a_{+}b_{+}} \right).
\label{eq:local_diag}
\end{equation}
As we discussed in Sec.~\ref{sec:model}, on the level of approximation taken through this work the relevant elements of the reduced density matrix are the diagonal ones, referred to the eigenbasis of $\hat{H}_\tf{sys}$. This restricts the Liouville space described here to the case where $a_{+}=a_{-}$ and $b_{+}=b_{-}$ and therefore $[W^A]^{bb}_{aa} = A_{aa}\delta_{ba}$. Employing the same notation as in the main text (i.e. representing the reduced density matrix as a vector) we obtain:
\begin{align}
D^{(i)}_{\nu \nu'} &= \bm{e}^\tf{T} \bm{W}^{F_\nu} \bar{\boldsymbol{\Pi}} \bm{W}^{F_{\nu'}} \bm{p}^{(i)} + (\nu \leftrightarrow \nu'), 
\label{eq:corr_fin} \\
\bar{\boldsymbol{\Pi}} &= \tilde{\bm{W}}^{-1}(\bm{p}^{(i)} \otimes \bm{e}^\tf{T} - \bm{1}), \nonumber
\end{align} 
where care must to be taken in not confusing $\bm{W}^{F_\nu}$ with the nonlocal $K$-kernels defined in Eq.~(\ref{eq:kernelA}).

\section{Recursive relation for the angular velocity}
\label{app:4}

In this appendix we give a recursive method to obtain the rotor's angular velocity. Starting from the 
angular Langevin equation Eq.~(\ref{eq:lang-theta}), we multiply both sides of the equation by $d\theta$ and 
integrate in the range $\theta_\tf{i} \leq \theta \leq \theta_\tf{f}$:
\begin{equation}
\dot{\theta}_\tf{f}^2 - \dot{\theta}_\tf{i}^2 = \frac{2}{\mathcal{I}} \int_{\theta_\tf{i}}^{\theta_\tf{f}}
\left( \mathcal{F}_\theta^{(i)} - \mathcal{F}_\tf{load} - \gamma_\theta \dot{\theta}  \right) d\theta,
\end{equation}
where we used that $\ddot{\theta} d\theta = \dot{\theta} d \dot{\theta}$ under the integral. We now 
consider as initial condition the unstable equilibrium point where $\mathcal{W}_\tf{eff}$ is 
maximum and take this angle as the origin, i.e. $\theta_\tf{i} = 0$. If the initial angular velocity is zero 
and we take $\theta_\tf{f} = \theta$, then the above equation can be written as
\begin{equation}
\dot{\theta}^2(\theta) = \frac{2}{\mathcal{I}} \int_0^\theta
\left( \mathcal{F}_{\theta'}^{(i)} - \mathcal{F}_\tf{load} - \gamma_{\theta'} \dot{\theta}(\theta') \right) d\theta'.
\end{equation}
The occurrence of $\dot{\theta}$ at both sides of the equation suggests the following functional recursion 
formula:
\begin{equation}
\dot{\theta}_{n+1}(\theta) = \sqrt{\frac{2}{\mathcal{I}} \int_0^\theta
\left( \mathcal{F}_{\theta'}^{(i)} - \mathcal{F}_\tf{load} - \gamma_{\theta'} \dot{\theta}_n(\theta') \right) d\theta'}.
\end{equation}
By choosing $\dot{\theta}_0(\theta) = 0$ as the initial case, we obtain the following first order approximation to $\dot\theta(\theta)$:
\begin{equation}
\dot{\theta}_1(\theta) = \sqrt{\frac{2}{\mathcal{I}} \int_0^\theta
\left( \mathcal{F}_{\theta'}^{(i)} - \mathcal{F}_\tf{load} \right) d\theta'}.
\end{equation}
As discussed in Sec.~\ref{subsec:regimes}, the operation condition for the motor is $\mathcal{W}_F - 
\mathcal{W}_\tf{load} \geq \mathcal{E}_\tf{dis}$. The maximum allowed loading work $\mathcal{W}_\tf{load}^*$ 
is the one from which the above inequality can no longer be fulfilled, and hence $\mathcal{W}_F - 
\mathcal{W}_\tf{load}^* = \mathcal{E}_\tf{dis}$. If we now approximate this condition through the $n$-order 
solution in the above recursive formula, we have
\begin{equation}
\mathcal{W}_F - \mathcal{W}_\tf{load}^* = \int_0^{2\pi} \gamma_\theta \dot{\theta}_n(\theta) d\theta.
\label{eq:recursive}
\end{equation}
For $n=0$ we obtain the trivial condition $\mathcal{W}_\tf{load}^* = \mathcal{W}_F$, meaning that the 
work done by the loading force needs to be smaller than that delivered by the motor, otherwise it gets stuck. 
This condition, however, does not take into account the dissipation. For $n=1$ we arrive to
\begin{equation}
\mathcal{W}_F - \mathcal{W}_\tf{load}^* = \int_0^{2\pi} \gamma_\theta \sqrt{\frac{2}{\mathcal{I}} \int_0^\theta
\left( \mathcal{F}_{\theta'}^{(i)} - \mathcal{F}_\tf{load}^* \right) d{\theta'}} d\theta,
\end{equation}
which coincides with Eq.~(\ref{eq:condition1}) and it is shown in dashed line in Fig.~\ref{fig:fig3}(b). 
Obviously, as $\mathcal{F}_\theta^{(i)}$ and $\gamma_\theta$ are in principle general functions of $\theta$, 
the above equation for $\mathcal{W}_\tf{load}^*$ needs to be solved numerically. Such a solution 
fits well the crossover between the ``operational'' and ``non-operational'' regimes for low biases. In the high bias regime, we need to take $n=4$ and $5$ in Eq.~(\ref{eq:recursive}) to reach convergence in trajectories A and B, respectively, as shown by the solid lines in Fig.~\ref{fig:fig3}(b).

\bibliographystyle{apsrev4-1_title}
\bibliography{cite}

\end{document}